\documentclass[sn-apa]{sn-jnl}% APA Reference Style 
\usepackage[table]{xcolor}
\usepackage{makecell}

\usepackage{graphicx}%
\usepackage{multirow}%
\usepackage{amsmath,amssymb,amsfonts}%
\usepackage{amsthm}%
\usepackage{mathrsfs}%
\usepackage[title]{appendix}%
\usepackage{xcolor}%
\usepackage{textcomp}%
\usepackage{manyfoot}%
\usepackage{booktabs}%
\usepackage{algorithm}%
\usepackage{algorithmicx}%
\usepackage{algpseudocode}%
\usepackage{listings}%
\usepackage{rotating}
\usepackage{cleveref}
\Crefname{figure}{Fig.}{Figs.}  % 大写：Fig. 1

\theoremstyle{thmstyleone}%
\theoremstyle{thmstyletwo}%

\theoremstyle{thmstylethree}%

\begin{document}

\title[Article Title]{Enhancing Academic Paper Recommendations Using Fine-Grained Knowledge Entities and Multifaceted Document Embeddings}


\author[1,2]{\fnm{Haixu} \sur{Xi}}\email{xihaixu@jsut.edn.cn}

\author[3]{\fnm{Heng} \sur{Zhang}}\email{zh\_heng@ccnu.edn.cn}

\author*[1]{\fnm{Chengzhi} \sur{Zhang}}\email{zhangcz@njust.edu.cn}

\affil[1]{\orgdiv{Department of Information Management}, \orgname{Nanjing University of Science \& Technology}, \orgaddress{\city{Nanjing}, \postcode{210094}, \country{China}}}
\affil[2]{\orgdiv{School of Computer Engineering}, \orgname{Jiangsu University of Technology}, \orgaddress{\city{Changzhou},\postcode{213001}, \country{China}}}
\affil[3]{\orgdiv{School of Information Management}, \orgname{Central China Normal University}, \orgaddress{\city{Wuhan},\postcode{430079}, \country{China}}}


\abstract{In the era of explosive growth in academic literature, the burden of literature review on scholars are increasing. Proactively recommending academic papers that align with scholars' literature needs in the research process has become one of the crucial pathways to enhance research efficiency and stimulate innovative thinking. Current academic paper recommendation systems primarily focus on broad and coarse-grained suggestions based on general topic or field similarities. While these systems effectively identify related literature, they fall short in addressing scholars' more specific and fine-grained needs, such as locating papers that utilize particular research methods, or tackle distinct research tasks within the same topic. To meet the diverse and specific literature needs of scholars in the research process, this paper proposes a novel academic paper recommendation method. This approach embeds multidimensional information by integrating new types of fine-grained knowledge entities, title and abstract of document, and citation data. Recommendations are then generated by calculating the similarity between combined paper vectors. The proposed recommendation method was evaluated using the STM-KG dataset, a knowledge graph that incorporates scientific concepts derived from papers across ten distinct domains. The experimental results indicate that our method outperforms baseline models, achieving an average precision of 27.3\% among the top 50 recommendations. This represents an improvement of 6.7\% over existing approaches.}




\maketitle

\section{Introduction}\label{sec1}

The rapid expansion of global scholarly output has greatly increased the volume of published research, intensifying the burden of literature review. Efficiently and accurately locating papers that align with a researcher’s interests and expertise has become a central challenge \citep{gusenbauer2020academic}. Addressing this challenge can facilitate the discovery of new research topics and the development of innovative methods. By proactively delivering relevant papers, academic recommendation systems can reduce search effort, a capability that has attracted substantial attention from both academia and industry \citep{kreutz2022scientific}.

Existing approaches typically emphasize either global content analysis \citep{zhu2021recommender} or collaborative filtering based on historical interactions \citep{murali2019collaborative}. Some methods further incorporate author–author relations \citep{du2020recommendation} or citation networks \citep{son2018academic} to assess similarity among papers or researchers. However, these techniques generally operate at a coarse level of granularity and underutilize fine-grained knowledge entities present in text \citep{cao2021dekr}. Consequently, they often cannot explain why a paper is recommended or how closely it matches a researcher’s specific interests—for example, particular tasks, questions, or methodological choices within the same topical area.

Recent studies have begun to consider fine-grained signals \citep{takahashi2022solutiontailor}, specific knowledge entities \citep{ni2020layered} , and knowledge graphs \citep{brack2021citation, cao2021dekr, imene2022knowledge} . Mainstream search engines likewise link queries to entities in knowledge bases to improve ranking. Despite these advances, current solutions still fall short of meeting diverse literature-review needs. Researchers frequently seek papers that address similar tasks yet adopt different methods or theoretical perspectives to build comprehensive views of a field. Existing techniques typically rely on static semantic relations in knowledge graphs and do not dynamically model semantic compositions among heterogeneous entity types conditioned on a given task, making it difficult to reveal diverse research perspectives and methodologies.

In summary, current recommendation methods insufficiently address the joint requirements of fine-grained relevance and diversity. To close this gap, a recommendation framework is proposed that models dynamic, diverse semantic combinations among heterogeneous fine-grained knowledge entities. The approach integrates multidimensional signals—titles, abstracts, citation relations, and fine-grained entities and their relations—into document embeddings, and computes similarity over concatenated vector representations to deliver fine-grained, content-based recommendations while enhancing diversity. Specifically, a fine-grained scientific knowledge graph (FG-SKG) is constructed to capture detailed entities—research tasks, methods, materials, and metrics—and their relations. This graph supports the creation of multidimensional embeddings that jointly encode core textual content, citation context, and automatically extracted knowledge entities, thereby aligning recommendations with researchers’ fine-grained information needs. Diversity is further promoted through controlled combinations and weighting schemes in the similarity computation.

This work makes two primary contributions. (i) A recommendation method is introduced that improves both coverage and accuracy by aligning results with researchers’ specific and diverse literature needs via dynamic composition of fine-grained entities. (ii) A systematic evaluation of embedding models and concatenation strategies is provided on the $STM-KG$ dataset \citep{brack2021coreference}. The proposed method achieves a Top-50 precision of 27.3\%, representing a 6.7 percentage-point improvement over baseline models. All datasets and source code are publicly available at: https://github.com/jsutxhx/SKG-AR. 

\section{Literature Review}\label{sec2}

This section reviews the state of the art in: (i) knowledge unit–based academic paper recommendation, (ii) graph-based recommendation for scholarly literature, and (iii) multidimensional information embedding in scientific documents.

\subsection{Academic Paper Recommendation Based on Knowledge Units}\label{s1subsec1}

Existing systems primarily recommend papers by identifying similar scholars or similar documents, using semantic representations of content \citep{ma2021chronological} or collaborative filtering over past interactions \citep{sakibcoll2020}. To mitigate data sparsity, many methods incorporate graph structure—e.g., social ties and citation networks \citep{hao2021paper, du2020recommendation, kanwal2024research}. However, the objectivity and neutrality of such graph-derived signals remain contested and may adversely affect recommendation diversity and fairness \citep{ekstrand2018all}. Meanwhile, comparatively less attention has been devoted to analyzing and mining knowledge-level content within papers \citep{wu2020comprehensive}, limiting support for fine-grained, content-specific recommendations.

\begin{table}[h]
\caption{\centering{Summary of representative studies of academic paper recommendation based on knowledge units.}}\label{tab:based_knowledge}%
\renewcommand{\arraystretch}{1.2} % Adjust row height for readability
\begin{tabular*}{\textwidth}{@{\extracolsep{\fill}}p{1.5cm}p{2cm}p{1.5cm}p{6.5cm}}
\toprule
Category & Author & Methods & Main findings \\
\midrule
Discourse-level & \citet{santosa2024s3par} & S3par & S3PaR, a sequential scientific paper recommendation system based on paper chapter structure, is put forth with the goal of supporting academic writing with context-related references. \\
\midrule
Sentence-level & \citet{ma2025citation} & Multi-task learning & A multi-task learning model based on a multilayer perceptron was developed to simultaneously perform citation recommendation and argumentative classification of citation sentences. \\
 & \citet{takahashi2022solutiontailor} & Vector \makecell[l]{similarities} & It recommends academic papers by calculating the difference between the vector similarities of research objective sentences and methodology sentences. \\
 \midrule
 Phrase-level & \citet{sarwar2021recommending} & Multi-level chronological learning-based & Using keyword extraction and cosine similarity calculation, suggest related articles from references that aid in understanding the content of a certain article. \\
 & \citet{mohamed2021academic} & Concept-based & Model researchers’ interests by extracting semantic concepts from articles they have authored, downloaded, or read. \\
 & \citet{luan2018information} & Scientific terms based & Builds a knowledge graph of scientific terms using semi-supervised natural language processing techniques, and it deduces latent links to suggest pertinent techniques and ideas for researchers. \\
\botrule
\end{tabular*}
\end{table}

With the advent of large-scale knowledge bases (e.g., DBpedia\footnote{Dbpedia (https://dbpedia.org/) aims to extract knowledge from Wikipedia texts and provide structured content.}, YAGO\footnote{YAGO (https://yago-knowledge.org/) is an open-source knowledge base automatically extracted from Wikipedia and other sources.} and Freebase\footnote{The Freebase API has now been replaced by the Google Knowledge Graph API, and Freebase.com was officially shut down in May 2016.}), mainstream search engines have begun linking queries to entities and ranking entity-centric results. As summarized in Table~\ref{tab:based_knowledge}, academic recommender research has likewise explored knowledge unit embeddings at multiple granularities: discourse-level segments (e.g., Introduction, Related Work; \citep{santosa2024s3par}), sentence-level units (e.g., research objective/method sentences; \citep{takahashi2022solutiontailor}), rhetorical types (objective/method/conclusion; \citep{ma2025citation}), phrase-level elements (keywords; \citep{sarwar2021recommending}), scientific concepts \citep{mohamed2021academic}, and task/method entities \citep{luan2018information}. Related efforts also target recommendations of specific fine-grained knowledge such as topics \citep{liu2021idea}, methods \citep{huang2013amrec}, and datasets \citep{viswanathan2023datafinder}.

These studies advance fine-grained relevance by recommending papers through knowledge-centric content features. Nevertheless, treating knowledge units in isolation risks ignoring their semantic structure and inter-unit relations, which constrains deeper understanding of document content and limits recommendation effectiveness. Further research is needed to model the structure and semantic relations among knowledge units so as to deliver recommendations that are more detailed, accurate, and diverse.

\subsection{Graph-based academic paper recommendation}\label{s2subsec2}
Graph-based recommendation methods enhance the matching of scholarly resources by explicitly modeling structured relations among papers and related entities. As summarized in Table~\ref{tab:graph-based}, prior work has leveraged author–author collaboration networks \citep{hwang2010coauthorship}, paper citation networks \citep{sakibcoll2020}, and combinations thereof \citep{kanwal2024research}). However, reliance on structure alone can introduce concerns about objectivity and fairness \citep{adomavicius_context-aware_2015}. To mitigate these issues, subsequent studies have integrated scholarly profiles (e.g., research interests and fields) with textual semantics, constructing more expressive heterogeneous knowledge graphs \citep{gupta2017scientific, du2020recommendation, hao2021paper}.

To meet the need for fine-grained, content-aware recommendations, the field has moved from holistic content modeling to fine-grained knowledge modeling. Recent approaches extract semantic units—knowledge entities, scientific concepts, or knowledge-bearing sentences—and assemble scientific knowledge unit graphs that fuse semantic information with structural relations \citep{ni2020layered, brack2021citation, cao2021dekr, imene2022knowledge, xi2024enhancing}. By embedding these graphs and performing structural reasoning, recommendation models more comprehensively capture scholar preferences, paper semantics, and latent associations, thereby improving both accuracy and interpretability of results.

\begin{table}[h]
\caption{\centering{Summary of representative studies of graph-based academic paper recommendation.}}\label{tab:graph-based}%
\renewcommand{\arraystretch}{1.2} % Adjust row height for readability
\begin{tabular*}{\textwidth}{@{\extracolsep{\fill}}p{1.5cm}p{2cm}p{1.5cm}p{6.5cm}}
\toprule
Category & Author & Methods & Main findings \\
\midrule
Relation network-based & \citet{hwang2010coauthorship} & co-authorship based & The co-authorship network of the academic community is generated, and then path exploration and screening are performed for specific tasks to identify recommended papers. \\
& \citet{sakibcoll2020} & Citation-based & Using public contextual information to improve similarity detection through 2-level citation relations. \\
& \citet{kanwal2024research} & RRMF & RRMF improves paper recommendations by integrating citation networks and author collaborations, using structural patterns and identifying key authors. \\
\midrule
Knowledge graph-based & \citet{hao2021paper} & Interests-based & A method leverages authors' evolving interests and academic graph structures to enhance paper recommendation accuracy and relevance. \\
 & \citet{du2020recommendation} & HNPR & HNPR uses heterogeneous networks with citation links, author collaborations, and research areas, employing random walks to assess paper relevance. \\
 & \citet{gupta2017scientific} & CCA & Their approach integrates semantic text embeddings and citation graph network embeddings into a shared space using Canonical Correlation Analysis (CCA). \\
 \midrule
 Fine-gained Knowledge graph-based & \citet{xi2024enhancing} & FG-SKG \& Multi-embedding & A multidimensional document embedding representation based on fine-grained knowledge entities and their relations. \\
 & \citet{brack2021citation} & Multi-embedding \& Concepts & Integrating document embeddings learned from textual content and citation graphs with the scientific concepts mentioned in the papers. \\
  & \citet{cao2021dekr} & GNN \& Deep network & A knowledge graph that connects datasets and method entities, and integrated graph structure and text feature representation through graph neural networks and deep collaborative filtering networks to achieve appropriate method recommendations for specific datasets. \\
\botrule
\end{tabular*}
\end{table}

Graph representations (entity and relation embeddings), combined with vector-space operations, link prediction, and meta-path reasoning, have been shown to enhance accuracy and, to some extent, diversity. Nonetheless, current multi-dimensional embedding and reasoning techniques still struggle to satisfy diverse literature needs—particularly when dynamic compositions of knowledge are required (e.g., assessing the current status of a task for a given research object or enumerating alternative methods for a specific task). Further research on multi-dimensional information embedding for academic papers is therefore warranted.

\subsection{Multidimensional information embedding method for academic papers}\label{s1subsec3}

Academic papers encode heterogeneous signals, including author identities and collaborations, citation links among papers, and fine-grained knowledge entities (e.g., research tasks, methods, datasets, and evaluation metrics) together with their semantic relations. A single embedding is typically insufficient to capture these multi-dimensional semantics. Multidimensional information embedding has therefore been adopted to model distinct semantic units separately and integrate them into a structured, information-rich document representation.

Representative approaches such as SPECTER \citep{cohan2020specter} and SPECTER2 \citep{singh2023scirepeval} leverage citation supervision with Transformer encoders to produce citation-aware paper embeddings. SciNCL \citep{ostendorff2022neighborhood} further introduces multi-view contrastive learning, constructing negatives from metadata and content similarity to enhance discriminability. Other methods operate at structural or semantic-role levels: HyperDoc2Vec \citep{han2018hyperdoc2vec} partitions documents into sections to derive hierarchical embeddings, whereas COSAEMB \citep{singh2024cosaemb} learns role-aware vectors from scientific discourse structure (e.g., Methods, Results, Discussion) to improve interpretability and modularity.

In this work, these strands are combined and extended with knowledge-entity–aware embeddings that explicitly encode tasks, methods, materials, and metrics. Document-level and citation-level vectors are integrated to form a unified, hybrid representation that fuses structure and semantics, thereby improving semantic coverage and representation fidelity for recommendation. Motivated by gaps in prior research, two questions are addressed: (i) can recommendations be tailored to scholars’ fine-grained intents at the level of knowledge entities? (ii) can the performance of paper recommendation systems be further improved by exploiting multidimensional embeddings of both knowledge graphs and documents?

\section{Methodology}\label{sec3}

The proposed recommendation framework comprises three stages. First, a fine-grained scientific knowledge graph is constructed and its nodes are embedded to represent tasks, methods, materials, and metrics. Second, document representations are composed by concatenating vectors along node–relation paths, yielding multidimensional embeddings that capture content at multiple granularities together with citation relations. Third, informed by observed literature-search behaviors, paper vectors are reweighted and concatenated to screen candidates and rank results. This pipeline enables candidate selection and re-ranking that address scholars’ diverse information needs while preserving fine-grained semantic specificity.

\subsection{Research Framework}\label{s2subsec1}

\begin{figure}[h]%
\centering
\includegraphics[width=0.9\textwidth]{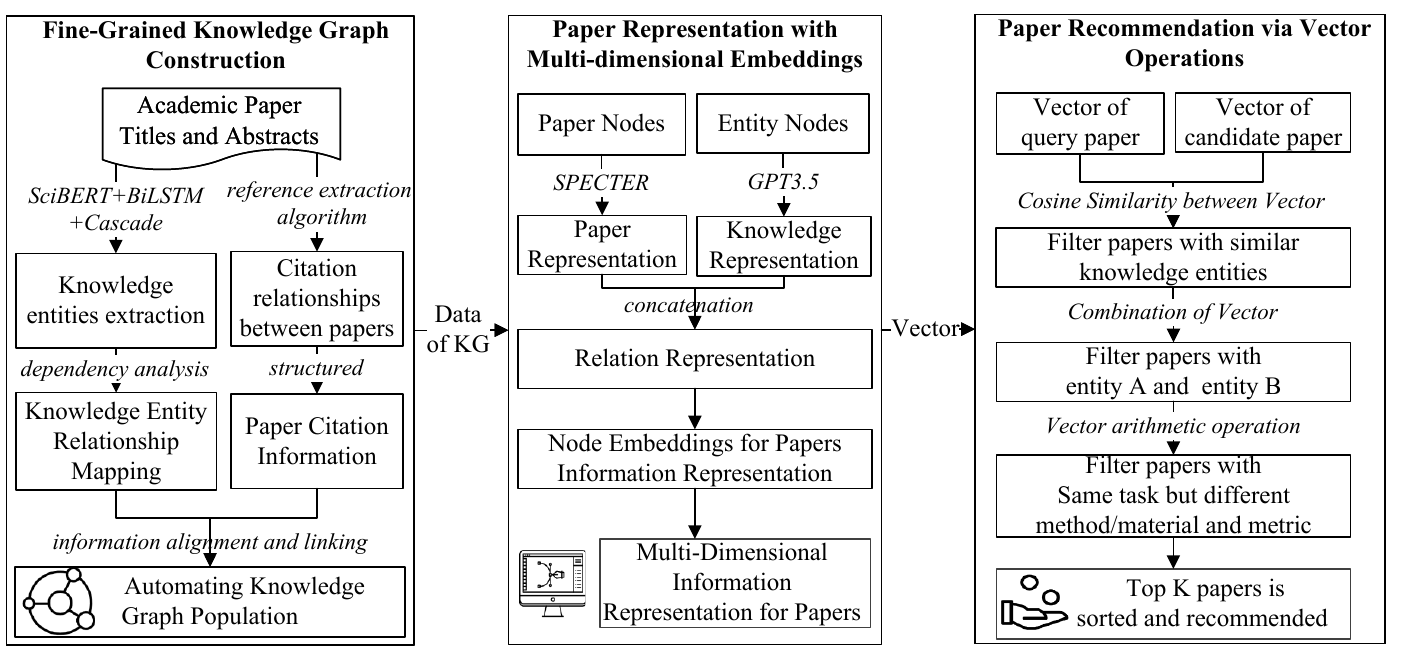}
\caption{\centering {Research Framework of paper recommendation.}}\label{fig1}
\end{figure}

\Cref{fig1} presents the overall architecture of the proposed method, which comprises three components: (i) construction of a fine-grained scientific knowledge graph (FG-SKG), (ii) multidimensional document embedding, and (iii) vector-based recommendation. By jointly leveraging the FG-SKG and multidimensional embeddings, the system is designed to balance precision and diversity, yielding recommendations that are both tailored and multifaceted.

The FG-SKG emphasizes detailed relational structure to enhance relevance. Four types of knowledge entities—tasks, methods, materials/datasets, and metrics—are extracted from titles and abstracts, and their relations are mapped to include both semantic links and citation-based connections, thereby enriching contextual signal. The resulting embeddings integrate textual content, citation information, and entity semantics to produce holistic vector representations of papers. During recommendation, candidates are first filtered via cosine similarity and then ranked with learned, behavior-informed weights, ensuring relevance while accommodating diverse scholarly information needs.

\subsection{Fine-grained Knowledge Graph Construction}\label{s3subsec2}

In this study, Fine-Grained Scientific Knowledge Graphs ($FG-SKGs$) are constructed by integrating research papers, their core knowledge entities, inter-entity relations, and inter-paper citation networks. The pipeline proceeds as follows. First, four categories of knowledge entities are identified from paper titles and abstracts. Next, intra-paper relations among these entities are extracted and typed. Then, inter-paper citation links are derived by analyzing bibliographic references. Finally, papers, entities, and their relations are fused into a structured, interconnected graph that provides a coherent, comprehensive view of the research landscape.

\subsubsection{Automatic Identification of Fine-gained Knowledge Entities in Academic Papers}\label{s1subsubsec1}

We first define the types of fine-grained knowledge entities relevant to scientific and technological papers and then develop an automated entity identification model whose performance is comparable to state-of-the-art (SOTA) methods.

(1)	Fine-grained knowledge entity categorization

Prior work has introduced a variety of entity types to structure scientific information \citep{zhang2024revealing, brack2021citation, hou2021tdmsci, luan2018multi}. including \textit{Task}, \textit{Method}, \textit{Tools}, \textit{Process}, \textit{Material}, \textit{Data}/\textit{Dataset}, \textit{Metric}, \textit{Score}, \textit{Other-ScientificTerm}, and \textit{Generic}. As the label set expands, identification accuracy typically declines, creating a trade-off between coverage and reliability in recommendation scenarios \citep{ni2020layered,brack2021coreference, dessi2022cs}. Among the above categories, \textit{Process} and \textit{Tools} capture method-specific details, and \textit{Score} represents concrete outcomes under \textit{Metric}; these elements are often not the primary focus during the early stages of literature exploration. \textit{Material}, \textit{Data}, and \textit{Dataset} refer to resources supporting research and can vary considerably across disciplines; for cross-domain consistency, they are unified here under \textit{Material}. Finally, \textit{Other-ScientificTerm} and \textit{Generic} lack sufficient specificity to capture nuanced user preferences. 

To balance specificity and robustness, four primary entity types are distilled: \textit{Task}, \textit{Method}, \textit{Material}, and \textit{Metric} (Table ~\ref{tab:entity_classify}). This categorization preserves the concepts most salient to scholarly search intent while controlling label complexity, thereby enhancing both the relevance and interpretability of recommendations.

\begin{table}[h]
\caption{\centering{Knowledge Entity Classification in Academic Papers.}}\label{tab:entity_classify}%
\renewcommand{\arraystretch}{1.2} % Adjust row height for readability
\begin{tabular*}{\textwidth}{@{\extracolsep{\fill}}p{1cm}p{1cm}p{4.2cm}p{5.6cm}}
\toprule
Our entities & Entities of exist & Entity Description & Example \\
\midrule
Task & Task & Describes the specific things or phenomena to be observed, analyzed, and studied, outlining the problems to be solved and the tasks to be completed, i.e., \textit{what is being researched}. & \setlength{\fboxsep}{1pt}%
\fcolorbox{white}{green}{Release information overload} of scholars is key problem in the era of mobile reading. The integration of syntactic steps and multi-feature text for automated structured \fcolorbox{white}{green}{abstracting of scientific papers} maybe the good choice. \\
Method & Method, Tools, Process & Explores the theoretical models, specific techniques, and procedures required to conduct research tasks, having proprietary names, i.e., \textit{how the research is conducted}. & \setlength{\fboxsep}{1pt}%
This paper proposes a \fcolorbox{white}{yellow}{Moves-Features summarization method} which integrates multiple text features into the iterative calculation process of the \fcolorbox{white}{yellow}{TextRank algorithm} by weight, introduces the \fcolorbox{white}{yellow}{MMR algorithm} for redundancy handling of the pre-selected abstract set. \\
Material & Material, Data and Dataset & Resources required to support research activities, including physical materials and digital corpora and datasets, i.e., \textit{what is used to conduct research}. & \setlength{\fboxsep}{1pt}%
Experimental studies on the \fcolorbox{white}{cyan}{SUMPUBMED dataset} show that compared to the baseline model, this method is efficient. \\
Metric & Metric & Parameters or indicators used to compare, measure, and evaluate the effectiveness of research outcomes, i.e., \textit{how the results of the research are evaluated}. & \setlength{\fboxsep}{1pt}%
This method has certain differences in improving \fcolorbox{white}{magenta}{ROUGE} for abstracts at different syntactic steps. \\
\botrule
\end{tabular*}

\vspace{0.5em}
\noindent\small\textbf{Note:} \quad Task: \colorbox{green}{\phantom{xx}}  \quad Method: \colorbox{yellow}{\phantom{xx}}  \quad Material: \colorbox{cyan}{\phantom{xx}}  \quad Metric: \colorbox{magenta}{\phantom{xx}}

\end{table}

(2)	Fine-grained knowledge entity identification model

\begin{figure}[h]%
\centering
\includegraphics[width=1.0\textwidth]{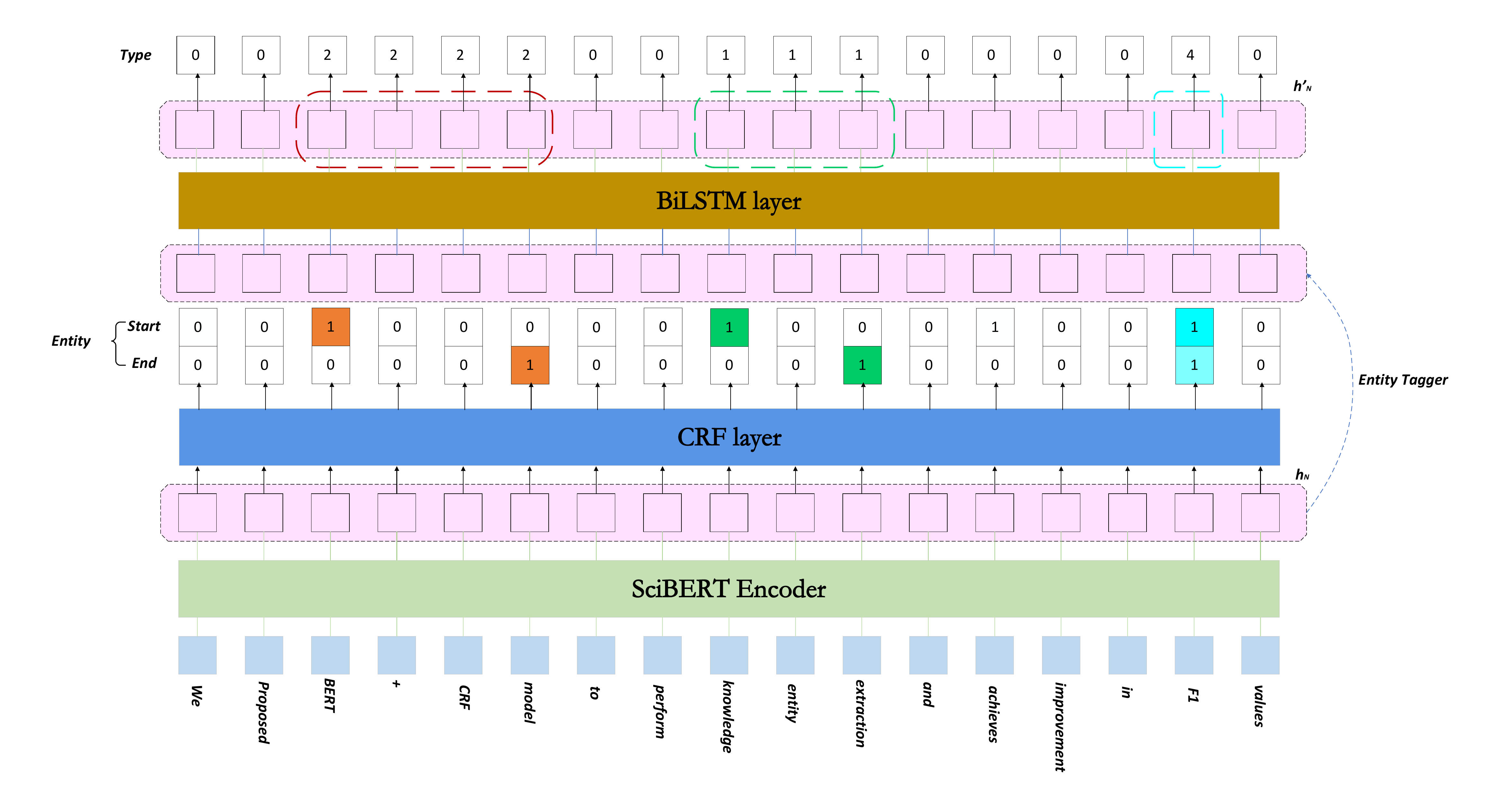}
\caption{\centering Deep Learning Models for Fine-Grained Knowledge Entity Identification.}\label{fig2}
\end{figure}

Accurate identification of fine-grained entities is essential for representing scholars’ detailed information needs. To this end, a \textit{SciBERT + BiLSTM + Cascade} architecture \citep{zhang2024revealing} is adopted. The model (\Cref{fig2}) incorporates data augmentation following Zhang et al. (2024) to increase sample size and diversity, improves token-level semantics via \textit{SciBERT}, captures contextual dependencies with a \textit{BiLSTM}, and performs cascaded learning for entity boundary detection and type assignment. A \textit{CRF} layer is employed to enforce valid label transitions and improve sequence consistency. 

Concretely, the pipeline comprises three components: (i) \textit{SciBERT} encoding, (ii) \textit{BiLSTM} contextualization, and (iii) cascade decoding (entity tagger + entity-type tagger). The input is first encoded into token vectors, yielding a hidden sequence H. In the cascade decoder, two binary classifiers identify start and end positions of entities; a \textit{CRF} then constrains the predicted tag sequence to ensure valid \textit{BIO}/\text{IOB} patterns, producing entity spans. Next, an entity-type classifier assigns one of the four target categories (\textit{Task}/\textit{Method}/\textit{Material}/\textit{Metric}) to each span. In parallel, the token-level representation H is further contextualized by the \textit{BiLSTM} to produce $H'$, from which span representations are derived and fed to the type classifier. This design improves multi-label classification efficiency while maintaining high boundary precision and robust type discrimination.

\subsubsection{Automatic Relations Extraction of Fine-gained Knowledge Entities in Academic Papers}\label{s1subsubsec2}

We first define the relation types among fine-grained knowledge entities relevant to scientific and technological papers. Based on these definitions, a type-to-relation mapping strategy is implemented to automatically infer relation types between entity pairs. The effectiveness of this extraction procedure is contingent on the accuracy of the underlying entity-type recognition model.

(1) Categorization of Fine-Grained Knowledge-Entity Relations

Recent studies have identified a wide range of relations among knowledge entities to support knowledge-graph construction \citep{dessi2022cs, brack2021citation, mondal2021end, luan2018multi}. These include \textit{Compare}, \textit{Conjunction}, \textit{Evaluate-for}, \textit{evaluatedOn}, \textit{evaluatedBy}, \textit{Part-of}, \textit{Feature-of}, \textit{Used-for}, \textit{HyponymOf}, \textit{Coreferent}, \textit{Related}, among others—amounting to 179 distinct semantic relations. During literature review, however, scholars typically seek works connected to specific entities rather than exploring every possible relational nuance. Accordingly, relations such as \textit{Compare}, \textit{Conjunction}, \textit{Part-of}, \textit{HyponymOf}, and \textit{Feature-of} can be consolidated into a broader \textit{Related} relation for retrieval purposes. In parallel, \textit{Used-for} is refined to capture type-specific interactions: \textit{achievedBy} (Task $\leftrightarrow$ Method), \textit{usedBy} (Material $\leftrightarrow$ Task), and \textit{evaluatedBy} (Task $\leftrightarrow$ Metric). These distinctions reflect the semantic roles of the four entity types and improve representational clarity.

To reduce extraction complexity without sacrificing coverage, explicit co-reference links are not modeled; instead, surface-form and semantic matching are applied against external knowledge bases (e.g., \textit{Wikidata}) to unify mentions. Consequently, a streamlined set of four essential relations—\textit{achievedBy}, \textit{usedBy}, \textit{evaluatedBy}, and \textit{related}—is proposed (Table ~\ref{tab:relation_entities}). To further capture nuanced same-type interactions, hierarchical refinements are introduced (Table ~\ref{tab:relation_entities}): \textit{Task} is subdivided into \textit{Object} and \textit{Problem}, with related linking \textit{Task}–\textit{Object} and \textit{Task}–\textit{Problem}; \textit{Method} is subdivided into \textit{Process}, with \textit{related} linking \textit{Method}–\textit{Process}.

\begin{table}[h]
\caption{\centering{Relations Among Knowledge Entities in Academic Papers.}}\label{tab:relation_entities}%
\renewcommand{\arraystretch}{1.2} % Adjust row height for readability
\begin{tabular*}{\textwidth}{@{\extracolsep{\fill}}p{1cm}p{1cm}p{6.5cm}p{4cm}}
\toprule
Primary Entity & Sub-Entity & Entity Description & Entity relation \\
\midrule
Task & Object & Specific things or phenomena directly observed, analyzed, and studied, defining the scope of research tasks, i.e., who is being studied. & Task \textit{related} Object \\
 & Problem & Subtasks or sub-problems within the research task. & Task \textit{related} Problem \\
Method & Process & Actions that change or generate information about the state of research objects in research methods. & \makecell[l]{ Method \textit{related} Process \\ Task \textit{achievedBy} Method } \\
Material & — & — & Material \textit{usedBy} Task \\
Metric & — & — & Task \textit{evaluatedBy} Metric \\
\botrule
\end{tabular*}
\end{table}

(2) Relation Extraction for Fine-Grained Knowledge Entities

Traditional pipelines detect relations after entity recognition, which can propagate errors and depress overall performance \citep{li2014incremental,shang2022onerel}. Motivated by the observation that many relations in abstracts are largely type-determined (Table ~\ref{tab:entity_classify}), relation extraction is framed as: (i) a binary decision of whether two entities are linked, followed by (ii) a type-to-relation mapping.

Because abstracts often contain multiple entities of the same type, a precision-oriented heuristic pairing strategy based on dependency syntax is adopted to avoid spurious pairings from static mappings. Each admissible type pair is associated with a small set of dependency-path templates (e.g., \textit{nsubj} + \textit{dobj}, \textit{nmod}, \textit{acl}; Table ~\ref{tab:dependency_path}). A relation is instantiated only if the entity pair (a) matches one of the templates and (b) exhibits locally coherent semantics; otherwise, no edge is produced. This design prioritizes high-confidence edges that are safe to inject into the KG and downstream recommendation.

A preliminary evaluation on annotated data indicates $Precision > 85\%$, $Recall \approx 60\%$, and $F1 \approx 0.70$, demonstrating practicality under low-annotation constraints. A small manual manual audit further confirms that template-matched relations are predominantly correct; these edges are therefore assigned higher weights during KG construction to mitigate error propagation from noisy edges and reduce recommender sensitivity to occasional omissions.

\begin{table}[h]
\caption{\centering{Dependency Path Template for Entity Relation Extraction.}}\label{tab:dependency_path}%
\renewcommand{\arraystretch}{1.2} % Adjust row height for readability
\begin{tabular*}{\textwidth}{@{\extracolsep{\fill}}p{3cm}p{2cm}p{3cm}p{1cm}p{1cm}p{1.5cm}}
\toprule
Entity combination & Relation & Typical Dependency path/ relation & Precision & Recall & $F_1-score$ \\
\midrule
Task—Method & achievedBy & nsubj+dobj, nmod & 0.85 & 0.65 & 0.73 \\
Task—Material & usedBy & nsubj+dobj, nmod & 0.88 & 0.62 & 0.73 \\
Task—Metric & evaluatedBy & nmod & 0.90 & 0.58 & 0.70 \\
Task—Object/Problem & related & nmod, acl & 0.87 & 0.63 & 0.73 \\
Method—Process & related & nsubj+dobj, nmod, acl & 0.83 & 0.60 & 0.70 \\
 
\botrule
\end{tabular*}
\end{table}

We later quantify the sensitivity to missing edges by relation type in §4.3.2 (Table \ref{tab:rel_type_impact}), showing that \textit{achievedBy} is most critical for accuracy (\textit{MAP}/\textit{nDCG}), whereas \textit{usedBy} contributes more to diversity (\textit{ILD}/\textit{Coverage}).”

It should be noted that the relation induction component is intended for large-scale scientific literature organization and recommendation, emphasizing scalability and low annotation cost; a fully supervised relation extractor is not trained at this stage. As dependency-based recall is limited by template coverage, future work will investigate joint modeling and semi-/weakly supervised relation extraction to improve coverage while retaining high precision.

\subsubsection{Automatic Population of Knowledge Graph of Academic Papers}\label{subsubsec3}

We first outline the structure of Fine-Grained Scientific Knowledge Graphs (FG-SKGs) relevant to scientific and technological papers and then describe an automated procedure for graph population. The effectiveness of FG-SKGs depends critically on the accuracy of entity recognition, since identifying entities and extracting their relations are prerequisite steps.

(1) Structure of the FG-SKGs

The FG-SKG is designed to systematically represent the essential elements of academic papers—abstracts, fine-grained knowledge entities, and relations both within and across papers. A heterogeneous information network formulation is adopted to flexibly encode multiple node and edge types in a unified, multi-relational space. This design supports efficient retrieval and traversal for recommendation tasks while capturing the inherent complexity of scholarly content (e.g., citation links and entity–entity semantics).

Formally, the knowledge graph is defined as: $G=(P,E,L_{PP},L_{PE},L_{EE})$, where $P$ denotes papers and $E$ denotes knowledge entities. Entities are categorized into four primary types: $E = T \cup M \cup D \cup R$, with $T$ (Tasks), $M$ (Methods), $D$ (Materials/Datasets), and $R$ (Metrics). For a paper $p \in P$, its entity set is $E_p = (T_p, M_p, D_p, R_p) \subseteq E$. Edges encode three relation families: 
(i) $L_{PP} \subseteq P \times P$ for citations between papers; 
(ii) $L_{PE} \subseteq P \times E$ for paper–entity links; and 
(iii) $L_{EE} \subseteq E \times E$ for entity–entity relations, including the four essential semantic links (e.g., \textit{achievedBy}, \textit{usedBy}, \textit{evaluatedBy}, \textit{related}) defined in Table~\ref{tab:relation_entities}. 
Each node/edge type stores both content attributes (e.g., abstracts, entity descriptions) and structural attributes (e.g., citation neighborhoods, typed semantic connections), yielding a dynamic, interconnected representation of research topics, methodologies, resources, and evaluation practices.

(2) Automated Population of FG-SKGs

To construct the knowledge graph, a multi-stage pipeline is employed to identify, classify, and link papers and their constituent entities. The process comprises:(i) entity recognition and intra-document relation induction, 
(ii) citation linkage, and 
(iii)graph materialization.

The model described in Section 3.2.1 is first applied to paper titles and abstracts to detect fine-grained entities—\textit{Task (T)}, \textit{Method (M)}, \textit{Material (D)}, and \textit{Metric (R)}—and to categorize them according to the predefined schema. Typed relations—\textit{achievedBy}, \textit{usedBy}, \textit{evaluatedBy}, and \textit{related}—are then induced as elements of $L_{EE}$. Extracted entities are stored as node sets, yielding a structured representation of each paper’s fine-grained content. 

In parallel, paper–paper citations ($L_{PP}$) and paper–entity links ($L_{PE}$) are established using citation-parsing tools and auxiliary scripts, enriching the KG with inter-paper and paper–entity connectivity.

The FG-SKG is materialized in either (a) a graph database supporting heterogeneous networks or (b) a single JSON file. Nodes and edges are formalized as follows:

\[
\text{Papers} = \left[ \{ \text{paper\_id}, \text{title}, \text{abstract} \}, \dots \right]
\]

\[
\text{Entities} = \left[ \{ \text{entity\_id}, \text{name}, \text{domains}, \text{top\_type} \}, \dots \right]
\]

\[
\text{data} = \left\{ 
\begin{aligned}
&\text{nodes}: \left[ \{ \text{id}, \text{category} \}, \dots \right], \\
&\text{links}: \left[ \{ \text{source}, \text{target}, \text{relation} \}, \dots \right], \\
&\text{categories}: \left[ \text{name}, \dots \right]
\end{aligned}
\right\}
\]

Here, nodes include papers and entities, each tagged with a category (paper, task, method, material, metric). Links encode typed relations, including $L_{PP}$ (citations), $L_{PE}$ (paper–entity associations), and $L_{EE}$ (entity–entity relations), and specify a source, target, and relation type. Categories enumerate node types to enable efficient filtering and query operations (e.g., task-focused or method-focused retrieval).

This structured design yields an effective, scalable FG-SKG that serves as the backbone for candidate generation and relation-aware re-ranking in the recommendation pipeline, leveraging interconnected entities and multi-hop relational paths to support precise and diversified scholarly recommendations.

\subsection{Embedding Representation of Papers}\label{s2subsec3}

A paper’s multidimensional information comprises content (titles/abstracts and fine-grained knowledge entities) and citation context. To satisfy diverse scholarly needs, these signals are embedded as vectors and combined according to node relations to form comparable representations at multiple granularities.

\subsubsection{Embedding of Graph Nodes}\label{s2subsubsec1}

The objective of embedding is to learn low-dimensional vectors such that papers that cite one another, share similar textual content, or contain similar fine-grained entities are close in the embedding space. Two node types are considered: papers and knowledge entities.

For paper nodes, the Transformer-based document encoder \textit{SPECTER} \citep{cohan2020specter} is employed. SPECTER jointly models title/abstract semantics and citation relations and has demonstrated strong performance in scholarly embedding tasks. Preprocessed titles and abstracts are fed to SPECTER to obtain a paper vector $\mathbf{s}_p$.

For entity nodes, a \textit{GPT-based} encoder is used to model short definitional or descriptive contexts. Because entity semantics are primarily sequential (rather than graph-structured) and typically lack citation links, \textit{GPT}’s sequence modeling is well suited. Directly applying \textit{SPECTER} to entities would introduce structural mismatch and require training a new entity-level corpus from scratch. Accordingly, entity descriptions are encoded with \textit{GPT} to obtain semantic vectors. Sub-entities under Task and Method are grouped consistently to maintain type coherence.

Entity embeddings within a paper are aggregated by sum pooling, a simple and interpretable set-representation technique with strong scalability and generalization \citep{zaheer2017deep}. Let $\mathbf{c}_p^t$, $\mathbf{c}_p^m$ and $\mathbf{c}_p^d$ denote the aggregated vectors for \textit{task}, \textit{method}, and \textit{material}/\textit{metric} entities, respectively, and let $GPT(\cdot)$ be the entity encoder:

\[
\mathbf{c}_p^t = \sum_{e_i^t \in T_p} GPT(e_i^t) \tag{1}
\]
\[
\mathbf{c}_p^m = \sum_{e_i^m \in M_p} GPT(e_i^m) \tag{2}
\]
\[
\mathbf{c}_p^d = \sum_{e_i^d \in D_p} GPT(e_i^d) \tag{3}
\]

\subsubsection{Vector Combination via Node Relations}\label{s2subsubsec2}

In practice, researchers query at different granularities (e.g., \textit{Task} only; \textit{Task} + \textit{Method}; \textit{Task} + \textit{Material}/\textit{Metric}; or all). To support such use cases prior to candidate selection and ranking, paper representations are composed by concatenating the paper vector with one or more aggregated entity vectors:

Overall representation (paper with all entities):
\begin{equation}
 \mathbf{p}_g = [\mathbf{c}_p^t, \mathbf{c}_p^m, \mathbf{c}_p^d, \mathbf{s}_p] \tag{4}
\end{equation}

Two-entity combinations:
\begin{equation}
    \mathbf{p}_{tm} = \left[ \mathbf{c}_p^t, \mathbf{c}_p^m, \mathbf{s}_p \right], \quad \mathbf{p}_{td}= \left[ \mathbf{c}_p^t, \mathbf{c}_p^d, \mathbf{s}_p \right].
    \tag{5}
\end{equation}

Single-entity combinations:
\begin{equation}
    \mathbf{p}_t = \left[ \mathbf{c}_p^t, \mathbf{s}_p \right], \quad \mathbf{p}_m = \left[ \mathbf{c}_p^m, \mathbf{s}_p \right], \quad \mathbf{p}_d = \left[ \mathbf{c}_p^d, \mathbf{s}_p \right].
    \tag{6} \label{eq:single}
\end{equation}

These combination provide multi-granular, comparable representations for downstream recommendation.

\subsection{Paper Recommendation via Vector Operations}\label{s1subsec4}

Scholars often pursue diverse information needs during literature search \citep{bawden2007turn}. Beyond finding topically similar papers, many intentionally consult studies that address the same task with different methods or materials to broaden perspective and foster innovation \citep{liu2019task, talley2011database}. After constructing multidimensional embeddings, candidate selection and ranking are formulated as similarity-based retrieval with learned combination weights.

(1) Design of Vector Operations

Given a query paper $q \in P$,
semantic similarities are computed along the fine-grained dimensions (\textit{task}, \textit{method}, \textit{material}/\textit{metric}) and for overall content. Let $cos_{task}$,$cos_{method}$, and $cos_{met\_metric}$ denote dimension-wise similarities and $cos_{enti}$ denote overall similarity. A target vector $\mathbf{p}$, is formed as a convex combination of four representations: overall $\mathbf{p}_g$, and single-entity combinations
$\mathbf{p}_t$, $\mathbf{p}_m$, $\mathbf{p}_d$:

\begin{equation}
\mathbf{p} \;=\; \sum_{i\in\{g,t,m,d\}} w_i\,\mathbf{p}_i,
\qquad
w_i \ge 0,\ \sum_{i} w_i = 1 .
\tag{7} \label{eq:overall}
\end{equation}

\begin{equation}
\mathrm{cos}(\mathbf{q},\mathbf{p})
\;=\;
\frac{\mathbf{q}^{\top}\mathbf{p}}{\lVert \mathbf{q}\rVert\,\lVert \mathbf{p}\rVert } .
\tag{8}
\end{equation}

Here, $p_g = [c_p^t, c_p^m, c_p^d, s_p]$; $p_t, p_m, p_d$ correspond to the respective single-entity combinations in Eq.~\eqref{eq:single}. Within the task-similar subset $CP_{task} \subset C_P$, multiple signals (same/different method; same/different material/metric) are aggregated by another learned convex combination:

\begin{equation}
\mathrm{cos}(\mathbf{q},\mathbf{p})
\;=\;
\sum_{j=1}^{4}\alpha_j\, s_j(\mathbf{q},\mathbf{p}),
\qquad
\alpha_j \ge 0,\ \sum_{j} \alpha_j = 1 ,
\tag{9} \label{eq:aggregate}
\end{equation}

where $\{s_j \}_{j=1}^4$ correspond, in order, to $cos_{task\_method}$ ($\mathbf{p}=\mathbf{p}_{tm}$), $cos_{task}-cos_{method}$, $cos_{task}-cos_{mat/metric}$, and $cos_{task\_mat/metric}$ ($\mathbf{p}=\mathbf{p}_{td}$). Candidate papers are ranked by:

\begin{equation}
\mathrm{rank}(\mathbf{q},\mathbf{p})
\;=\;
\sum_{i\in\{g,t,m,d\}} w_i\,\mathrm{cos}(\mathbf{q},\mathbf{p}_i)
\quad\text{or}\quad
\sum_{j=1}^{4}\alpha_j\, s_j(\mathbf{q},\mathbf{p}) .
\tag{10}
\end{equation}

and the $top-N$ items are returned.

(2) Learning and Optimizing Combination Weights

The weights in Eqs.~\eqref{eq:overall} and ~\eqref{eq:aggregate} are learned on a development split, with accuracy-tuned (such as \textit{MAP}) as the sole objective ($\lambda=0$) and then fixed for testing. A two-stage procedure is used: 
(i) coarse grid search (step 0.05) initialized by the original heuristic profile; 
(ii) coordinate ascent with grid refinement (step 0.01), projecting updates back to the simplex with early stopping after five non-improving iterations. 
A sensitivity analysis perturbs one weight at a time by $\pm 10\%$ and $\pm 20\%$ (followed by renormalization), with all other settings held constant. 
On STM-KG (in-domain), each configuration is run with three random seeds; mean $\pm$ standard deviation is reported, and significance is assessed via paired $t$-tests.

\begin{figure}[h]%
\centering
\includegraphics[width=0.9\textwidth]{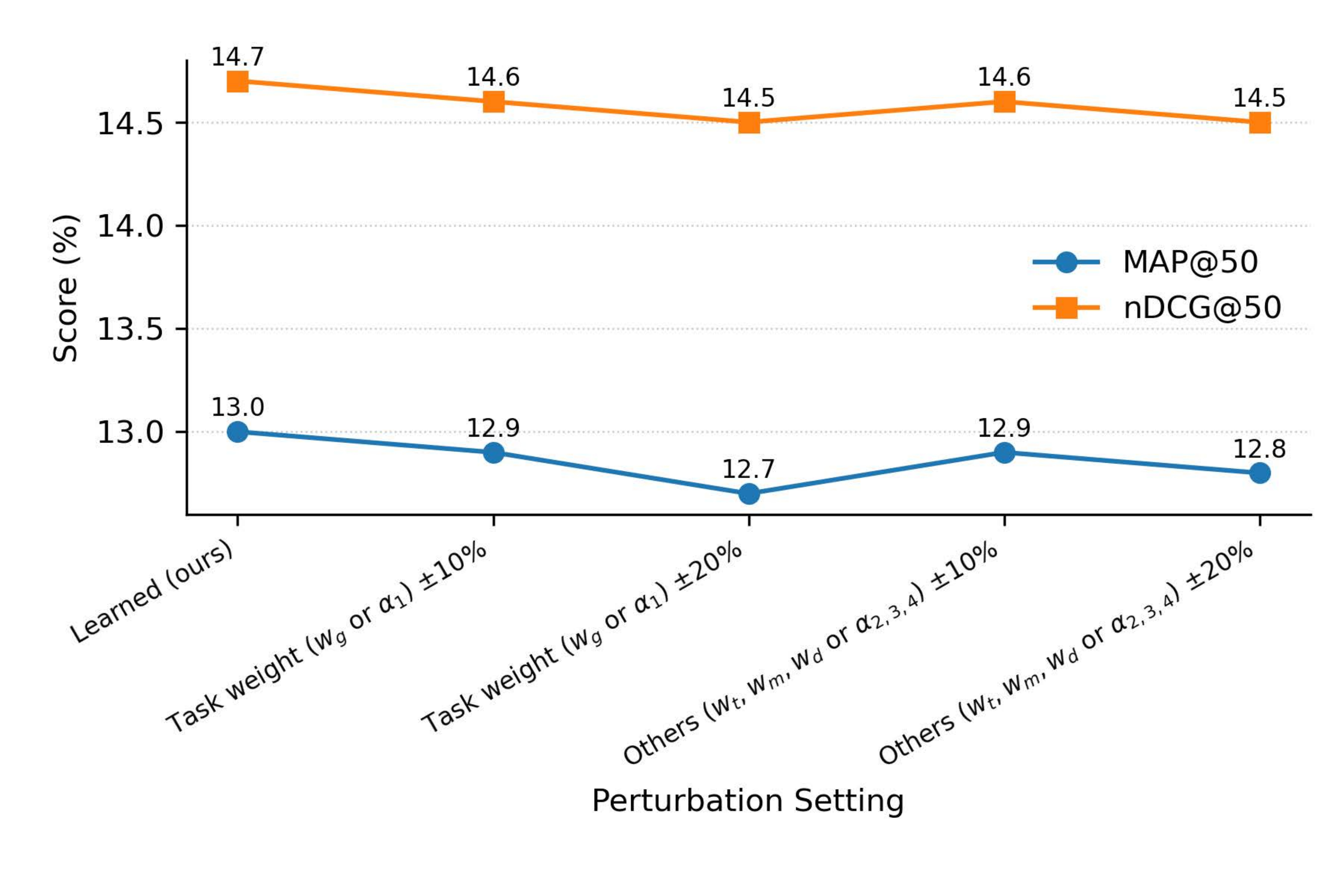}
\caption{\centering Sensitivity to Weight Perturbations on STM-KG (In-domain).}\label{fig3}
\end{figure}

As illustrated in \Cref{fig3}, the learned weights yield stable and significant improvements in both MAP@50 and nDCG@50, while local perturbations around the optimum induce performance changes $\leq 1.5\%$, indicating robustness. 
Reducing the task weight leads to the largest drop, confirming the dominant role of task semantics in scholarly similarity; moderate perturbations to method and material/metric weights have smaller effects, suggesting complementary contributions. The learned weights on STM-KG (in-domain) are: 

\[
w^\star = (0.46, 0.28, 0.18, 0.08) \quad \text{and} \quad \alpha^\star = (0.40, 0.30, 0.20, 0.10),
\]

with item order matching the original formulation. Gains are reproduced on an independent test set.

\subsection{Algorithmic Details of Paper Recommendation}\label{s1subsec5}

To improve reproducibility and clarity, we summarize the complete pipeline from FG-SKG construction to ranking. The steps correspond to §3.2–§3.4 and Eqs. (1)–(10): node embeddings (§3.3.1), multi-vector composition (§3.3.2), candidate generation and task-aware refinement (§3.4), and weight learning with sensitivity analysis.

\subsubsection{Pseudocode of Our Method}\label{s1subsubsec1}

Appendix B show the pseudocode for the recommendation flow proposed by this paper. Table~\ref{tab:symbol_algthim} list the symbols and definition used in the pseudocode. Cosine scoring in Eq. (8) over a candidate pool of size $|\mathbf{CP}| = K$ with vector dimension $d$ costs $O(K \cdot d)$. 
Task-aware refinement in Eq. (9) adds another $O(K \cdot d)$. 
Concatenation in Eqs. (4)–(6) is performed once per paper ($O(d)$). 
Relation extraction (\S 3.2) is run offline and does not affect test-time complexity.

\begin{table}[t]
\caption{\centering{Symbols and definitions used in §3.3–§3.4 Extraction.}}\label{tab:symbol_algthim}%
\renewcommand{\arraystretch}{1.1}
\begin{tabular}{ll}
\toprule
Symbol & Definition \\
\midrule
$P$, $E$ & Set of papers; set of entities in FG-SKG \\
$T_p,M_p,D_p$ & Task, Method, Material/Metric entities of paper $p$ \\
$\mathbf{s}_p$ & SPECTER embedding of paper $p$ (title+abstract) \\
$\mathrm{GPT}(e)$ & GPT-based embedding of entity $e$ \\
$\mathbf{c}_p^t,\mathbf{c}_p^m,\mathbf{c}_p^d$ & Sum-pooled entity vectors (Task/Method/Material) for $p$ \\
$\mathbf{p}_g,\mathbf{p}_t,\mathbf{p}_m,\mathbf{p}_d$ & Concatenated paper vectors (Eqs. 4–6) \\
$\mathbf{q}$, $\mathbf{p}$ & Query and target paper vectors \\
$w_i$ & Nonnegative weights over $\{g,t,m,d\}$ with $\sum_i w_i=1$ (Eq. 7) \\
$\alpha_j$ & Nonnegative weights over four task-subset signals with $\sum_j \alpha_j=1$ (Eq. 9) \\
$CP$, $CP_{\text{task}}$ & Candidate pool; task-similar subset \\
$\cos(\cdot,\cdot)$ & Cosine similarity (Eq. 8) \\
$\mathrm{rank}(\mathbf{q},\mathbf{p})$ & Final ranking score (Eq. 10) \\
\bottomrule
\end{tabular}
\end{table}

\subsubsection{Differences from Prior Work}\label{s1subsubsec1}

(1) Against single-vector text encoders. Prior systems typically represent each paper with a single topic-level vector, limiting control and interpretability. Our approach composes multiple entity-aware vectors (Task/Method/Material/Metric) and learns convex weights (Eqs. 7 and 10), enabling fine-grained control over which aspects drive similarity and reasons of recommendations.

(2) Against pure citation-based recommenders. Citation-based methods can overemphasize popularity and risk fairness or diversity issues. We fuse text semantics, entity semantics and citation context, and our precision-first relation induction (§3.2) injects high-confidence edges, improving ranking stability and robustness to missing links (see §4.4.2).

(3) Against knowledge-unit methods without dynamic composition. Existing unit-based methods often treat units in isolation or rely on static roles. We introduce task-aware refinement (Eq. 9) that recombines signals conditionally (same/different method/material) to satisfy common scholarly intents ("same task, different method/material").

The main reason why our method is superior to above mentioned methods is that: (i) Learned convex weights yield statistically significant gains over uniform weighting and single-vector baselines, while sensitivity analysis shows $\leq$1.5\% fluctuation under ±10\%–±20\% perturbations; (ii) precision-oriented relation induction avoids error propagation from noisy edges, providing consistent net benefits; (iii) multi-vector composition delivers interpretable, tunable, and diversity-aware recommendations aligned with fine-grained user needs.

\section{Result Analysis}\label{sec4}

In order to validate the effectiveness of the paper recommendation method based on fine-grained knowledge entities, this study extracted fine-grained knowledge entities from the \textit{STM-KG} \citep{brack2021coreference} dataset to construct a fine-grained knowledge graph. Subsequently, academic paper recommendation based on the combination of multidimensional document embeddings was implemented. The experimental results of both the knowledge entity extraction and academic paper recommendation subtasks were then compared and analyzed against relevant benchmarks.

\subsection{Dataset}\label{s3subsec1}

This paper draws on the series of studies by Brack et al. and uses the STM‑corpus dataset to verify the knowledge entity task and the STM-KG dataset to verify the paper recommendation task. The \textit{STM-corpus} \citep{brack2020domain} contains 110 paper abstracts, each belonging to 10 different scientific fields (agriculture, astronomy, biology, chemistry, computer science, earth science, engineering, materials science, mathematics, medicine), and 11 abstracts in each field. The \textit{STM-KG} \citep{brack2021coreference} is a knowledge graph constructed based on 55,485 academic paper abstracts in the above 10 disciplines, formed by automatically extracting scientific concepts. The graph covers paper titles, abstracts, and citation networks (including 15,395 citation links, including 2,200 interdisciplinary links), and annotates fine-grained knowledge entities and their relations. This structure is highly consistent with the modeling goal of "multidimensional semantic combination and entity enhanced representation" in our recommendation model. This dataset is used for the citation recommendation subtask based on the knowledge graph. It has good annotation quality and reuse basis, which provides strong support for us to evaluate different entity embedding strategies and recommendation performance. Although \textit{STM-KG} is less well-known than some general datasets (such as \textit{OpenAlex} and \textit{AMiner}), its advantages in structural completeness and entity semantic information make it very suitable for verifying our proposed recommendation method.

Since the types of knowledge entities used in this study do not entirely align with the types of scientific concepts in this graph, papers containing complete title and abstract information from \textit{STM-corpus} dataset are selected as entity recognition corpus. After sentence segmentation, the \textit{BRAT} annotation tool was employed. Building upon the abstract annotations of 110 papers in the \textit{STM Corpus} \citep{brack2020domain}, annotations for entities such as \textit{object}, \textit{task}, \textit{problem} entities were added and a model for automatic identification of entities was retrained. We use the \textit{BIO} annotation format to identify entity boundaries and formulate a unified annotation specification (as shown in Appendix A) to clarify the definition standards, context discrimination strategies, nested or overlapping conflict handling methods of various entities. The annotation was completed by 3 graduate students with NLP or computer science backgrounds. All annotators received training before formal annotation and passed pre-annotation tests to ensure consistent understanding. Each sample was completed independently by two annotators, and the conflicting samples were reviewed by an experienced researcher. We randomly selected 300 samples from the annotation set and calculated the Cohen’s Kappa value at the entity level. The average consistency score was 0.82, indicating that the annotations were highly consistent. The statistics of entity annotation results are shown in Table~\ref{tab:anno_result}.

To prevent training–test entanglement through contemporaneous citation links, we enforce a chronological split between encoder pretraining and our evaluation window. Concretely, all evaluation papers and their incoming/outgoing citations post-date the encoder’s pretraining cutoff; the encoder is frozen and no citation edges from the evaluation period are used to fit embedding parameters in our pipeline. This protocol eliminates the possibility that evaluation-time citation relations leak into representation learning.

Subsequently, we selected papers with at least 4 citations from the remaining \textit{STM-KG} as query papers, and constructed a recommendation evaluation set containing 720 papers and 4069 citation links. After constructing the knowledge entity graph of these papers using the fine-grained knowledge entity recognition model trained in Section 3.2.1 and the entity relation mapping method in Section 3.2.2, various embedding vectors are used to represent each query paper, and vector operations are used to retrieve the Top $K$ papers from all other papers in the dataset for recommendation.

\begin{table}[h]
\centering
\caption{\centering Entity annotation and data argument results of \textit{STM-corpus} dataset.}
\label{tab:anno_result}
\renewcommand{\arraystretch}{1.3}
\begin{tabular*}{\textwidth}{@{\extracolsep{\fill}}p{3cm}p{1cm}p{1cm}p{1cm}p{1cm}p{1cm}p{1cm}p{1cm}}
\toprule
\textbf{Category} & \textbf{Task} & \textbf{Object} & \textbf{Problem} & \textbf{Method} & \textbf{Process} & \textbf{Material} & \textbf{Metric} \\
\midrule
Quantity of Entity annotation &
811 & 698 & 287 & 258 &
1,301 & 2,099 & 1,658 \\
Quantity of Entity argument &
1,072 & 923 & 379 & 341 &
1,720 & 2,774 & 2,191 \\
Quantity of Sentence & \multicolumn{7}{l}{Total: 2,487} \\
Quantity of Sentence augment & \multicolumn{7}{l}{Total: 3,287; Training: 2,629; Validation: 329; Test: 329} \\
\bottomrule
\end{tabular*}
\end{table}

\subsection{Evaluation Metrics}\label{s1subsec2}

The evaluation metrics of this study include both knowledge entity extraction and paper recommendation subtasks. For knowledge entity extraction, \textit{Precision}, \textit{Recall}, and $F_1-score$ are commonly used metrics. We use span-level strict match to evaluate knowledge entity recognition, requiring that the start and end positions of the predicted entity must be exactly the same as the annotated entity, the boundaries must be exactly the same, and the entity category must be the same to be considered a "hit"; otherwise, regardless of the degree of partial overlap, it is considered a "miss". This judgment setting is consistent with mainstream named entity recognition evaluation standards such as CoNLL-2003 \citep{tjong2003introduction}, which helps to accurately reflect the model's boundary judgment ability.

The purpose of paper recommendation is to suggest diverse papers that align with scholars' specific research needs. In this study, we evaluate the accuracy of the recommendation method by measuring the overlap between the references of the query paper and the recommended results that have a similarity greater than the average similarity. A higher overlap indicates that the recommended papers better match the scholar's needs. Additionally, we assess the diversity of the recommendations by comparing the overlap between the references of the query paper and the recommended results. A lower overlap suggests that the recommendation method offers scholars a more diverse range of papers for their specific research tasks.

Both overlap values are calculated using the Mean Average Precision at K ($MAP@K$) metric. $MAP@K$ is the mean of the average precision ($AP@N$) of the top N recommended results across all query papers. As shown in equation (11), $AP@N$ measures the precision of the top 1 to N recommended results for a specific query paper in terms of how accurately they align with the target papers.

\[
AP@N(p) = \frac{\sum_{k=1}^N \text{Precision@}k(p) \cdot \text{rel}(k)}{\# \text{ relevant documents for } p} \tag{11}
\]

$Precision@k$ is the proportion of the first $k$ recommendation results that are cited. If the recommended result at position $k$ is cited, $rel(k)$ takes the value 1, otherwise it takes the value 0.

In addition to considerning whether relevant literature is recommended, $nDCG@K$ Further focus on the ranking of relevant literature in the recommended list, emphasizing that highly relevant literature should appear as high as possible. As shown in equation (12)

\[
\text{\textit{nDCG}@}K(p) = \frac{\sum_{i=1}^{K} \frac{2^{r_i} - 1}{\log_2(i + 1)}}{\text{\textit{IDCG}@}K(p)} \tag{12}
\]

$IDCG@K(p)$ representing the $DCG$ value under ideal sorting, used for normalization. $rel(i)$ represents the relevance level of the $i-th$ position in the recommendation result, usually taken as 0 (irrelevant) or 1 (relevant).

We introduce two diversity-oriented indicators and mainly apply to our method with using learned weights for combining different vector components:
\begin{itemize}
\item Intra-List Diversity ($ILD$): Determines the average pairwise dissimilarity of recommended papers. A higher $ILD$ score suggests greater variation in the $top-K$ list. As shown in equation (13)

\[
\mathrm{\textit{ILD}@}K \;=\; \frac{2}{K(K-1)} \sum_{1 \le i < j \le K} \Big(1-\cos\!\big(f(r_i),f(r_j)\big)\Big) \tag{13}
\]

$f(\cdot)$ denote the embedding function used for diversity evaluation.

\item Item Coverage: The proportion of unique papers among the $top-K$ recommendations for all test queries. This statistic measures the recommendation system's breadth and uniqueness at the corpus level. As shown in equation (14)
\end{itemize}

\[
\mathrm{Coverage@}K \;=\;
\frac{\big|\{\,b(r)\,|\, r \in R_K\,\}\big|}     {\big|\mathcal{B}_{\mathrm{cand}}\big|} \tag{14}
\quad
\]

$\mathcal{B}_{\mathrm{cand}}$ be the set of semantic buckets over entities spanned by the candidate pool for this query, $b(r) \in B$ maps a paper to its bucket and $R_K=\{r_1,\dots,r_K\}$ be the top-$K$ recommendation list for a query paper.

\subsection{Experimental Results of Knowledge Entity and Relation Extraction}\label{s3subsec3}
\subsubsection{Experimental Setup}\label{s3subsubsec1}

In order to avoid training bias caused by differences in the length of literature and ensure that each subset contains all entities, this paper randomly divides the labeled dataset into training set, validation set, and test set at the sentence level, with a ratio of 8:1:1. We compared the performance of knowledge entity recognition using \textit{BiLSTM+CRF} (referred to as Method a), \textit{SciBERT} (referred to as Method b), \textit{SciBERT+CRF} (referred to as Method c), and the approach proposed in this paper.

To ensure the stability and representativeness of the model evaluation results, we performed a 9-fold cross validation on the remaining 90\% of the sentence data while keeping the original test set (10\%) unchanged. We divided this part of the data into 9 equal parts, each fold was used as a validation set in turn, and the rest was used for training, and the final evaluation was performed on the fixed test set.

To further verify whether the performance difference between the proposed method and the baseline model is significant, we conducted a statistical test on the $F_1-score$ of the \textit{SciBERT} and \textit{SciBERT+BiLSTM+Cascade} models in the 9-fold cross validation.

To evaluate the effectiveness of Large Language Models (LLMs) in academic knowledge entity recognition, we compared supervised models with zero sample and few sample entity extraction methods based on Current mainstream LLM: \textit{GPT3.5}. We used title and abstract from the \textit{STM-corpus} as input for \textit{GPT-3.5} through \textit{OpenAI} API. Each text is independently processed and compared to the gold annotations. We tested two \textit{LLM} prompt strategies:

(1) Zero samples: Require the model to extract entities from the abstract using only natural language instructions (without using any labeled examples).

(2) Few samples (3 examples): Add three labeled examples before the prompt to guide the model in context learning.

To evaluate the contribution of high-precision relations extracted from dependency syntactic templates to recommendation performance and their robustness to missing relations, we conducted two controlled experiments on the $STM-KG$ dataset. The first experiment focused on contribution evaluation: while maintaining consistent recommendation model parameters, we compared the $Entity-only$ model (knowledge entities only) with the $Entity+Rel$ model (which adds high-confidence relations verified by dependency path template matching and small-scale manual spot checks). The second experiment focused on missing relations robustness: using the $Entity+Rel$ model as the baseline, we randomly retained a certain proportion of relation edges,$r \in \{100\%, 75\%, 50\%, 25\%\}$,
while keeping all other training and evaluation settings unchanged to simulate further recall deficiency. The two sets of experiments were independently run three times in-domain. The mean and standard deviation of $MAP@50$ and $nDCG@50$ is reported, and the significance is evaluated by paired t-test.

As shown in Table~\ref{tab:ie_hparams}, We summarize the hyperparameters only for the information extraction components, relation induction is rule-based and requires no trainingas.

\begin{table}[t]
\centering
\caption{Key hyperparameters for knowledge entity recognition, Runtime: Python~3.11.5.}
\label{tab:ie_hparams}
\renewcommand{\arraystretch}{1.12}
\begin{tabular}{lcccccc}
\toprule
\textbf{Module} & \textbf{Method} & \textbf{Batch} & \textbf{LR} & \textbf{Hidden Dim} & \textbf{Epochs} & \textbf{Max Seq} \\
\midrule
Entity recognition & Method~a & 64  & 0.001  & 128 & --- & --- \\
 & Method~b & 512 & 2e-5   & --- & --- & 512 \\
 & Method~c & 512 & 2e-5   & --- & 4   & 512 \\
\multicolumn{2}{l}{Relation extraction} & \multicolumn{5}{l}{\makecell[l]{uses dependency-path templates (heuristic),\\ hence no trainable hyperparameters.}}\\
\bottomrule
\end{tabular}
\end{table}

\subsubsection{Experimental Results and Analysis}\label{s3subsubsec2}

(1) Comparative analysis of entity indentification

The \textit{Precision}, \textit{Recall}, and $F_1-score$ of knowledge entity identification under different models are shown in Table~\ref{tab:enti_inden_result}. The proposed \textit{SciBERT} + \textit{BiLSTM} + \textit{Cascade} model, exhibited the best performance with an micro $F_1-score$ of 85.10. This superior performance can be attributed to several factors. Firstly, compared to Method a, this study utilized the \textit{SciBERT} pre-trained language model, specifically tailored for scientific papers, which provides highly discriminative representations of sentence-level semantic features. Additionally, data augmentation techniques were employed to expand the originally labeled corpus of 110 papers, increasing the diversity and quantity of training samples, thus enhancing the model's robustness. Furthermore, the Cascade method facilitated cascading learning for entity recognition and classification tasks, reducing cascade errors. In comparison to Method b and Method c, this approach not only mitigated cascade errors but also provided a larger training dataset. Notably, the experimental results revealed a slight negative impact on knowledge entity extraction when adding the \textit{CRF} layer after the \textit{SciBERT} layer. A detailed analysis of the extraction results suggested that the entity order learned by \textit{CRF} interfered with the contextual relations captured by the \textit{SciBERT} model. While the \textit{SciBERT} model accurately embeds word semantics, it also captures contextual relations, obviating the need for additional \textit{BiLSTM} layers to achieve satisfactory knowledge entity extraction results.

From the $F_1-score$ of different models on each entity type, the proposed \textit{SciBERT+BiLSTM+Cascade} remains best on \textit{Task}/\textit{Method}, while the \textit{SciBERT} and \textit{SciBERT+CRF} models have a slight lead in the $F_1-score$ on the \textit{Material} and \textit{Metric} entities, respectively. \textit{SciBERT} captures goal/problem-oriented semantics in task sentences and domain-specific phrases in method sentences; \textit{BiLSTM} smoothes long clauses and dependencies; and the \textit{Cascade} stage first detects entities and then types them, eliminating entity type confusion and reducing error propagation.
Material-related entity phrases can be sparse and have unique spellings, and pure \textit{SciBERT}'s domain-specific pre-trained subword statistics already fit them well, while the addition of \textit{BiLSTM}/\textit{Cascade} may over-regularize or bias compound phrases, introducing slight bias. Due to \textit{CRF}'s stronger label transition modeling capabilities and boundary consistency, they have better recognition capabilities for shorter phrase labels, such as \textit{Metric} entities. \textit{CRF}'s label transition constraints reduce boundary flipping around numbers/symbols and adjacent units, resulting in sharper boundaries and fewer BIO errors.

As shown in Table~\ref{tab:enti_inden_result}, the results indicate that the \textit{GPT3.5} model exhibits strong generalization ability, especially in low sample environments. However, its overall performance still lags behind our supervised model. In addition, the output of \textit{GPT3.5} sometimes suffers from structural inconsistency and entity boundary overlap, which limits their reliability in downstream tasks such as knowledge graph construction and paper recommendation. The fine-tuned supervised model proposed in this paper demonstrates excellent accuracy, category separation, and output consistency in fine-grained academic environments.

As shown in Table~\ref{tab:9-fold_result}, the main model \textit{(SciBERT+BiLSTM+Cascade)} performs stably on different test subsets, with the $F_1-score$ ranges from -0.50 to +0.39, with very small fluctuations ($\sigma = 0.28$, full range is 0.89), verifying the rationality of the current data partitioning and the robustness of the method.The p-value is 0.004 using the paired sample t-test and 0.0039 using the Wilcoxon signed rank test. Both tests indicate that the performance difference is statistically significant (p $<$ 0.05), indicating that the proposed model has a stable and reliable performance advantage in the fine-grained entity recognition task.

\begin{table}[h]
\centering
\caption{Per-class Results of fine-grained Knowledge entity Identification under Different Models.}\label{tab:enti_inden_result}
\renewcommand{\arraystretch}{1.2} % Adjust row height for readability
\begin{tabular*}{\textwidth}{@{\extracolsep{\fill}}p{2.5cm}p{2cm}p{1cm}p{1cm}p{1.5cm}}
\toprule
\textbf{Model} & \textbf{Entity Type} & \textbf{Precision} & \textbf{Recall} & \textbf{$F_1$-score} \\ 
\midrule
\multirow{5}{*}{BiLSTM + CRF} & Task    & 78.50  & 80.20  & 79.34  \\
& Method& 80.60  & 81.10  & 80.85 \\
& Material& 78.10  & 79.40  & 78.74  \\
& Metric  & 80.3 & 82.00  & 81.14  \\
& \textit{Micro-average} & \textit{79.36} & \textit{80.85} & \textit{80.10} \\
\midrule
\multirow{5}{*}{SciBERT} 
 & Task     & 83.10 & 84.50 & 83.79 \\
 & Method   & 84.00 & 85.30 & 84.65 \\
 & Material & 84.50 & 83.00 & \textbf{83.74} \\
 & Metric   & 83.90 & 85.00 & 84.45 \\
 & \textit{Micro-average} & \textit{83.78} & \textit{84.77} & \textit{84.27} \\
\midrule
\multirow{5}{*}{SciBERT+CRF} 
 & Task     & 82.30 & 85.10 & 83.68 \\
 & Method   & 82.70 & 86.40 & 84.51 \\
 & Material & 83.00 & 84.20 & 83.60 \\
 & Metric   & 84.90 & 85.70 & \textbf{85.30} \\
 & \textit{Micro-average} & \textit{82.12} & \textit{85.83} & \textit{83.93} \\
\midrule
\multirow{5}{*}{\makecell{SciBERT+BiLSTM\\+Cascade (Proposed)}}
 & Task     & 85.90 & 85.60 & \textbf{85.75} \\
 & Method   & 86.40 & 86.00 & \textbf{86.20} \\
 & Material & 84.20 & 82.70 & 83.44 \\
 & Metric   & 83.70 & 84.30 & 84.00 \\
 & \textit{Micro-average} & \textit{84.84} & \textit{85.36} & \textit{\textbf{85.10}} \\
\midrule
\multirow{5}{*}{\makecell{9-fold cross validation\\ (average)}} 
 & Task     & 85.10 & 84.80 & 84.95 \\
 & Method   & 85.50 & 84.90 & 85.20 \\
 & Material & 84.00 & 83.20 & 83.60 \\
 & Metric   & 84.40 & 84.10 & 84.25 \\
 & \textit{Micro-average} & \textit{85.04} & \textit{84.80} & \textit{84.92} \\
\midrule
\multirow{5}{*}{GPT-3.5 Zero-shot} 
 & Task     & 70.40 & 72.60 & 71.48 \\
 & Method   & 72.00 & 73.90 & 72.94 \\
 & Material & 70.10 & 73.00 & 71.52 \\
 & Metric   & 72.50 & 74.80 & 73.63 \\
 & \textit{Micro-average} & \textit{71.25} & \textit{73.58} & \textit{72.39} \\
\midrule
\multirow{5}{*}{\makecell{GPT-3.5 Few-shot\\ (3 examples)}} 
 & Task     & 76.10 & 78.90 & 77.47 \\
 & Method   & 77.80 & 79.90 & 78.84 \\
 & Material & 75.50 & 79.20 & 77.31 \\
 & Metric   & 77.30 & 79.30 & 78.29 \\
 & \textit{Micro-average} & \textit{76.92} & \textit{79.34} & \textit{78.11} \\
\bottomrule
\end{tabular*}
\end{table}

\begin{table}[h]
\centering
\caption{\centering {Results of 9-fold cross validation on SciBERT and Our method.}}\label{tab:9-fold_result}
\centering
\renewcommand{\arraystretch}{1.2} % Adjust row height for readability
\begin{tabular*}{\textwidth}{@{\extracolsep{\fill}}p{1.5cm}p{3cm}p{5cm}}
\toprule
\textbf{Fold} & \textbf{SciBERT}\_${F_1}$ & \textbf{(SciBERT+BiLSTM+Cascade)}\_${F_1}$ \\
\midrule
Fold 1                 & 84.49              & 84.79\\
Fold 2                 & 84.23              & 85.00\\
Fold 3                 & 83.81              & 85.05\\
Fold 4                 & 84.31              & 84.42\\
Fold 5                 & 84.56              & 84.81\\
Fold 6                 & 84.20              & 85.19\\
Fold 7                 & 84.50              & 84.62\\
Fold 8                 & 84.37              & 85.31\\
Fold 9                 & 84.73              & 85.08\\
\botrule
\end{tabular*}
\end{table}

To further evaluate the entity identification performance of our model on other datasets. We collected two relevant open datasets, namely \textit{SciERC} and \textit{TDM}, to compare the evaluation results of our model with existing models. Table~\ref{tab:enti_inden_diff_data} shows that our model achieved the highest $F_1-score$ of 69.77 on the \textit{TDM} dataset and performed well on the \textit{SciERC} dataset. The result demonstrates the excellent generalization ability of our method for scientific entity identification.

\begin{table}[h]
\centering
\caption{\centering \textbf{Evaluation of entity identification model on \textit{SciERC} and \textit{TDM} dataset.}}
\label{tab:enti_inden_diff_data}
\renewcommand{\arraystretch}{1.2}
\begin{tabular*}{\textwidth}{@{\extracolsep{\fill}}p{1cm}p{3cm}p{3.7cm}p{1.2cm}p{1cm}p{1.5cm}}
\toprule
\textbf{Dataset} & \textbf{Author} & \textbf{Model} & \textbf{Precision} & \textbf{Recall} & \textbf{F\textsubscript{1}-score} \\
\midrule
\multirow{4}{*}{SciERC} 
& \citep{zhong2021frustratingly} & PURE & - & - & 68.90 \\
& \citep{eberts2020span} & SpERT & 70.87 & 69.79 & 70.33 \\
& \citep{zaratiana2022hierarchical} & Hierarchical Transformer & 67.99 & 74.11 & \textbf{70.91} \\
& Our & SciBERT+BiLSTM+Cascade & 66.95 & 71.49 & 69.14 \\
\midrule
\multirow{3}{*}{TDM} 
& \citep{hou2019identification} & SCIIE & 67.17 & 58.27 & 62.40 \\
& \citep{zaratiana2022hierarchical} & Hierarchical Transformer & 65.56 & 70.21 & 67.81 \\
& Our & SciBERT+BiLSTM+Cascade & 68.84 & 70.73 & \textbf{69.77} \\
\bottomrule
\end{tabular*}
\end{table}

(2) Relation-type-specific missing analysis

On an annotated sample, we estimate omission rates and typical error sources for each relation type; representative failure cases are listed to facilitate replication.

\begin{itemize}
    \item \textit{achievedBy} (Task–Method) — omission rate $\approx$ 6\%. The reason is that long-distance dependencies (task and method split across clauses), nominalized or elided method mentions, and coordination that breaks the dependency chain. For example: ``We address \textless Task\textgreater\ by a novel \textless Method\textgreater\ \dots'' where the method is split across sentences, preventing a valid path match;
    
    \item \textit{usedBy} (Material–Task) — omission rate $\approx$ 10\%. The reason is that prepositional or appositive realizations (``using/with/based on''), and generic placeholders (``a standard benchmark'') masking specific datasets/materials. For example: ``We evaluate \textless Task\textgreater\ on the \textless Dataset\textgreater\ \dots'' parsed as a modifier rather than a link;
    
    \item \textit{evaluatedBy} (Task–Metric) — omission rate $\approx$ 3\%. The reason is that implicit metric mentions (``macro-F1 of 0.91'' lacking an explicit ``F1'' token), symbol/number adjacency causing BIO boundary shifts. For example: ``achieving macro-averaged precision and recall \dots'' not consolidated under a unified metric entity;
    
    \item \textit{related} (intra-type) — omission rate $\approx$ 21\%. The reason is that hierarchical or synonymous variants distributed across sentences; cross-sentence coreference not materialized as explicit edges (we instead rely on external KB matching).  
\end{itemize}

Most misses occur in the broad \textit{related} class (often dispersed phrasings and long-distance dependencies), followed by \textit{usedBy} (varied lexical realizations of resource use). \textit{achievedBy} is comparatively well captured due to strong predicate cues; \textit{evaluatedBy} benefits from stable evaluation verbs and templates. Relation induction uses high-precision dependency templates; recall is known to be template-coverage-limited.

(3) Analysis of contribution of entity relation

\begin{figure}[h]%
\centering
\includegraphics[width=0.9\textwidth]{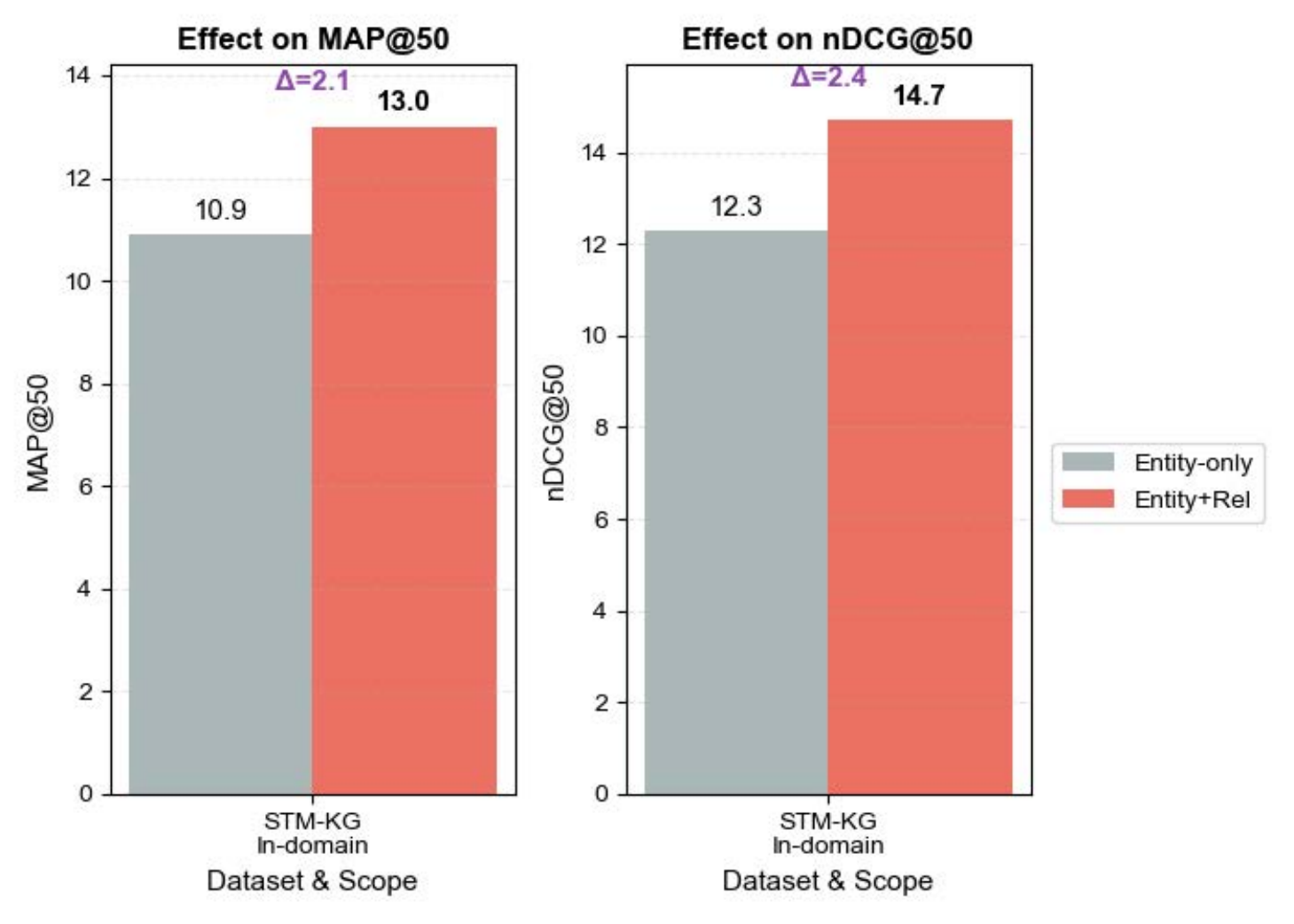}
\caption{\centering Contribution of High-Precision Relations in STM-KG (In-domain).}\label{fig4}
\end{figure}

In terms of the contribution of high-precision knowledge entity relationships, as shown in \Cref{fig4}, $Entity+Rel$ achieves stable and significantly improved recommendation performance on STM-KG (in-domain) compared to $Entity-only$: both $MAP@50$ and $nDCG@50$ improve by approximately +2.1 and +2.4 absolute points (paired t-test, p $<$ 0.01)respectively. This demonstrates that high-confidence relationships provide structural constraints for knowledge entities that complement the text, effectively improving ranking quality.

\begin{figure}[h]%
\centering
\includegraphics[width=0.9\textwidth]{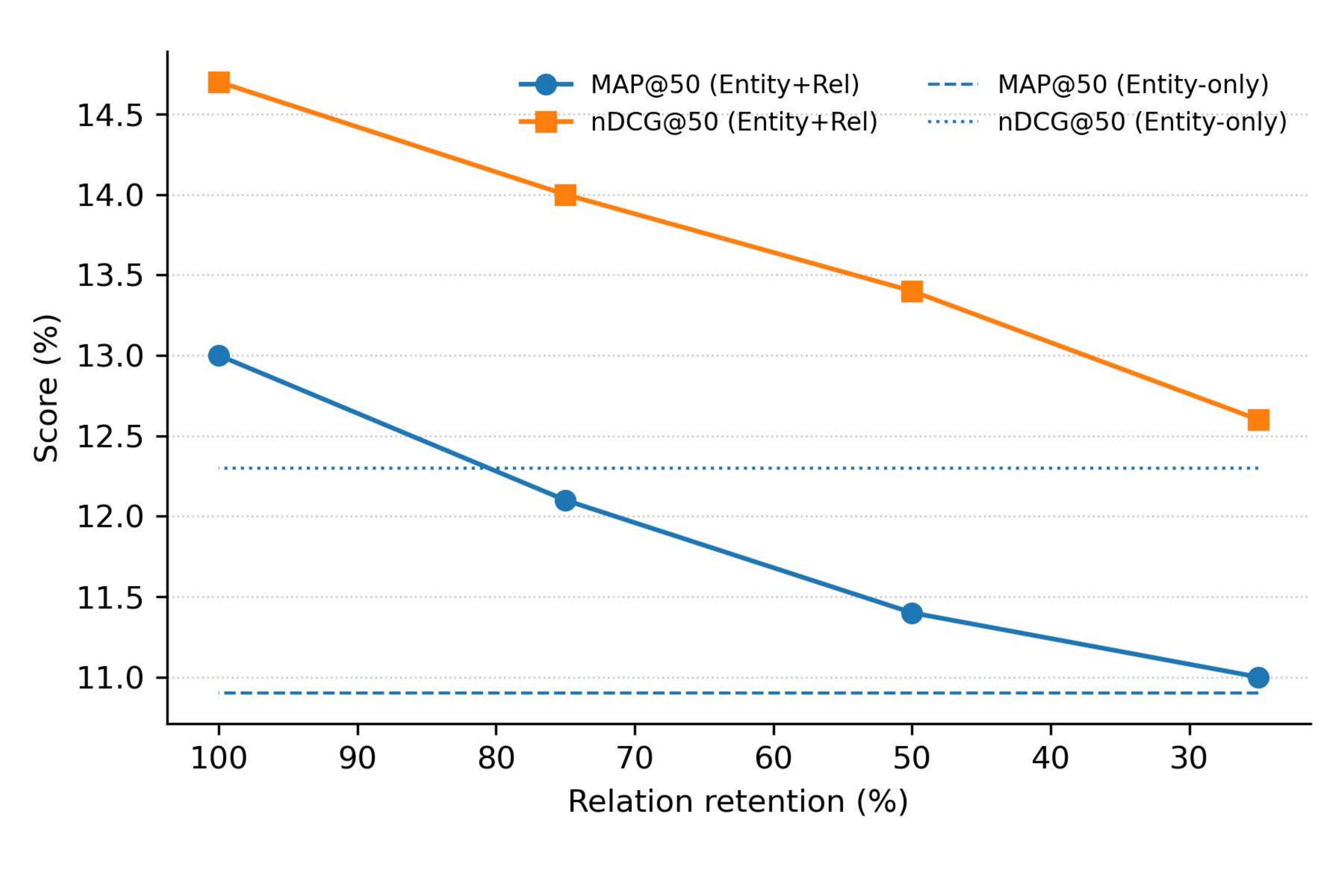}
\caption{\centering Robustness to Missing Relations in STM-KG (In-domain).}\label{fig5}
\end{figure}

In terms of robustness to missing relations, as shown in \Cref{fig5}, while keeping all other settings unchanged, the retention ratio of available relation edges is gradually reduced from 100\% to 75\% to 50\% to 25\%. Both metrics show a nearly linear and uniform decline: at 50\% retention, $MAP@50$ / $nDCG@50$ still maintains a significant positive gain over entity-only (approximately +0.5 to +1.1, $p < 0.01$); even at 25\% retention, performance remains comparable to entity-only (approximately +0.1 to +0.3, $p \approx 0.05$). These results demonstrate that the relations extracted using dependency templates are usable and effective, and the quality of their correct edges is sufficient to provide a net benefit. Furthermore, the system exhibits manageable performance degradation and sustained structural contribution to missing relations, validating our design trade-off of prioritizing high precision and gradually expanding recall.

(4) Relation-type-specific impact on recommendation

Table \ref{tab:rel_type_impact} shows that When removing \textit{achievedBy}, the accuracy (\textit{MAP}, \textit{nDCG}) decreases the most, but the diversity decreases relatively mildly, indicating that the "task-method" link mainly contributes to precise matching signals. Conversely, when removing \textit{usedBy}, the \textit{ILD} and \textit{Coverage} decrease significantly, indicating that the resource channel is the main source of result diversity. The impact of \textit{evaluatedBy} and \textit{related} is moderate or minor, more reflected in fine-grained rearrangement and slight contraction of coverage boundaries.

\begin{table}[t]
\centering
\caption{Impact of removing each relation type on recommendation metrics (STM-KG, in-domain; $\Delta$ vs. full graph). Negative values denote drops. Mean$\pm$std over 3 runs; paired $t$-tests for significance.}
\label{tab:rel_type_impact}
\renewcommand{\arraystretch}{1.12}
\begin{tabular}{lcccc}
\toprule
\textbf{Removed relation type} & $\Delta$MAP@50 & $\Delta$nDCG@50 & $\Delta$ILD@50 & $\Delta$Coverage@50 \\
\midrule
achievedBy (Task--Method) & -1.6 & -1.4 & -1.2 & -1.0 \\
usedBy (Material--Task)   & -1.0 & -0.9 & -3.8 & -3.4 \\
evaluatedBy (Task--Metric)& -0.6 & -0.5 & -1.5 & -1.2 \\
related (intra-type)      & -0.4 & -0.3 & -0.8 & -0.6 \\
\bottomrule
\end{tabular}
\vspace{2pt}
\small
Notes: Each row removes only one relation type while keeping others intact. Data splits, candidate pools, and evaluation follow §§4.1–4.4 to ensure comparability.
\end{table}

Although this strategy of relation extraction used by this paper has good practicality in applications, it has certain limitations when dealing with complex context-sensitive relations, and has not been directly compared with the existing supervised learning relation extraction model. In future work, it is planned to explore joint modeling or semi-supervised relation extraction methods to further improve the relation extraction effect, and on this basis, conduct a systematic comparative analysis with mainstream relation extraction models.

\subsection{Experimental Results of Academic Paper Recommendation}\label{s2subsec4}
\subsubsection{Experimental Setup}\label{s4subsubsec1}

Our proposed method uses \textit{SPECTER} embedding to represent the paper title, abstract and citation information, uses \textit{GPT3.5} embedding to represent the knowledge entities extracted from the paper title and abstract and use learned weights for combining different vector components to achieve paper recommendation. This experimental setup mainly includes comparative analysis of models on \textit{STM-KG} and Aminer\footnote{\url{https://www.aminer.org/citation}}  v12  dataset, and experiments of ablation study on \textit{STM-KG} dataset.

(1) Comparative analysis of models

In the baseline comparison experiment, our method without using learned weights for combining different vector components is compared with two similar methods, namely Baseline 1 \citep{takahashi2022solutiontailor}, Baseline 2 \citep{brack2021citation}  and three state-of-the-art models, namely SMIGNN \citep{niu2024smignn}, TIDRec \citep{xiao2025tidrec}, LLM-enhanced \citep{liu2025academic} in both in-domain and cross-domain paper recommendation scenarios under \textit{STM-KG} and Aminer v12 dataset. This experiment aims to verify the superior performance of our approach in paper recommendation accuracy and to assess the robustness and generalization ability of each approach when applied across domains.

\begin{itemize}
    \item Baseline 1: the method which utilizes the semantic similarity difference between research objective and method sentences; 
    \item Baseline 2: the method which builds paper embeddings from scientific concepts extracted from abstracts and \textit{SPECTER} embedding represents the paper title, abstract and citation information.
    \item SMIGNN: Node attention mechanism and intention attention mechanism are introduced to extract writing-citation graph and author graph to obtain the author's research interests.
    \item TIDRec: Use a main model that integrates dual-graph knowledge to capture the author's global research interests, and use a single-graph knowledge auxiliary model to explore local research interests. Triple-graph interactive distillation is used to correct the global research interests, and finally recommend highly relevant papers to the author.
    \item LLM-enhanced: LLM-enhanced citation network model for improved paper representation using large language models.
\end{itemize}

(2) Experiments of ablation study

To evaluate the impact of vector combination weights, node text embedding models, different types of knowledge entities, and citation information on our model, we conducted a set of ablation study. 
\begin{itemize} 
\item We investigate the impact of different word embedding models on in-domain paper recommendation’s accuracy. Specifically, we apply three embedding models—FastText \citep{bojanowski2017enriching}, \textit{SciBERT}, and \textit{GPT-3.5} \citep{radford2018improving}—to obtain semantic representations of document content. We compute the average of word embeddings over the corresponding textual units (title, abstract or entities) and use these document-level vectors for similarity-based recommendation.
\item We investigate the impact of different knowledge entity types and citation information on in-domain paper recommendation’s accuracy. Specifically, we compare the performance of paper recommendation using different types of document information embeddings after removing the material/metric, method, and task entities, as well as citation information. If citation information is ignored, the paper title and abstract are represented by the \textit{GPT3.5} model as one item of the document information embeddings. 
\item We investigate the impact of weighted combination strategies on in-domain and cross-domain paper recommendation performance. Specifically, under the same setting using \textit{GPT-3.5} and \textit{SPECTER} for semantic representation, we compare our method with and without the use of learned weights for combining different vector components. This allows us to reveal the influence of the introduction of weighting schemes on in-domain paper recommendations and cross-domain interdisciplinary paper recommendations.
\item We investigate orthogonal text-only baselines and ablations under the temporal decontamination protocol:
(i) BM25 (text-only); (ii) SciBERT (text-only); (iii) Our (text-only): replace the citation-informed encoder with a text-only encoder while keeping the rest of our pipeline; (iv) Our (full): with text and citation. Additionally, we ablate citation-side structure by removing co-citation and bibliographic-coupling features. All settings share the same candidate pool, training recipe, and evaluation metrics ($MAP@K$, $nDCG@K$, and an entity-aware relevance measure). We run 3 seeds and report mean and std; significance is assessed by paired t-tests.
\item To assess whether highly cited papers amplifies citation features, we report stratified results by citation tertiles of the query set (Low/Med/High) and also evaluate on a balanced subset that caps per-paper citation counts (top-coded at the 80th percentile). 
\end{itemize}

As the paper recommendation stage is a non-trainable vector retrieval and weighted re-ranking pipeline. For reproducibility, we also list the weight-search settings (objective, search schedule, and sensitivity analysis) and the vector dimensions used for document and entity embeddings, as shown in Table~\ref{tab:weight_search}.

\begin{table}[t]
\centering
\caption{Settings for weight learning / re-ranking (non-trainable) and vector dimensions.}
\label{tab:weight_search}
\renewcommand{\arraystretch}{1.12}
\begin{tabular}{@{}p{4.4cm}p{10.0cm}@{}}
\toprule
\textbf{Item} & \textbf{Setting} \\
\midrule
Vector dimensions & SPECTER paper embedding: \textbf{768}; GPT-3.5 entity embedding: \textbf{1536}. Concatenated vectors follow definitions $p_g=[c_t,c_m,c_d,s_p]$, $p_t=[c_t,s_p]$, $p_m=[c_m,s_p]$, $p_d=[c_d,s_p]$. \\
Objective (accuracy-tuned) & MAP@50 on dev split, $\lambda=0$; learned weights fixed for test-time ranking. \\
Objective (diversity-aware, optional) & $J_{\text{div}}=\text{MAP@}50+\lambda\cdot\text{ILD@}50$ (used only for trade-off analysis in §4.4.3). \\
Search space \& constraints & Simplex: $w_i\ge0,~\sum_i w_i=1$; $\alpha_j\ge0,~\sum_j \alpha_j=1$. \\
Search schedule & Coarse grid: step 0.05; coordinate ascent + fine grid: step 0.01; project back to simplex each step; early stop after 5 non-improving iters (dev MAP@50). \\
Sensitivity analysis & One-dim perturbations of $\pm10\%$ and $\pm20\%$ (renormalized), 3 seeds, paired $t$-tests. \\
Implementation notes & Cosine similarity; $\ell_2$-norm for vectors; temporal split to avoid leakage. \\
\bottomrule
\end{tabular}
\end{table}

\subsubsection{Experimental Results and Analysis}\label{s4subsubsec2}

We performed each experiment in the experimental setup on the specified datasets, recorded the experimental results and analyzed the possible reasons for this result.

\begin{figure}[h]%
\centering
\includegraphics[width=0.9\textwidth]{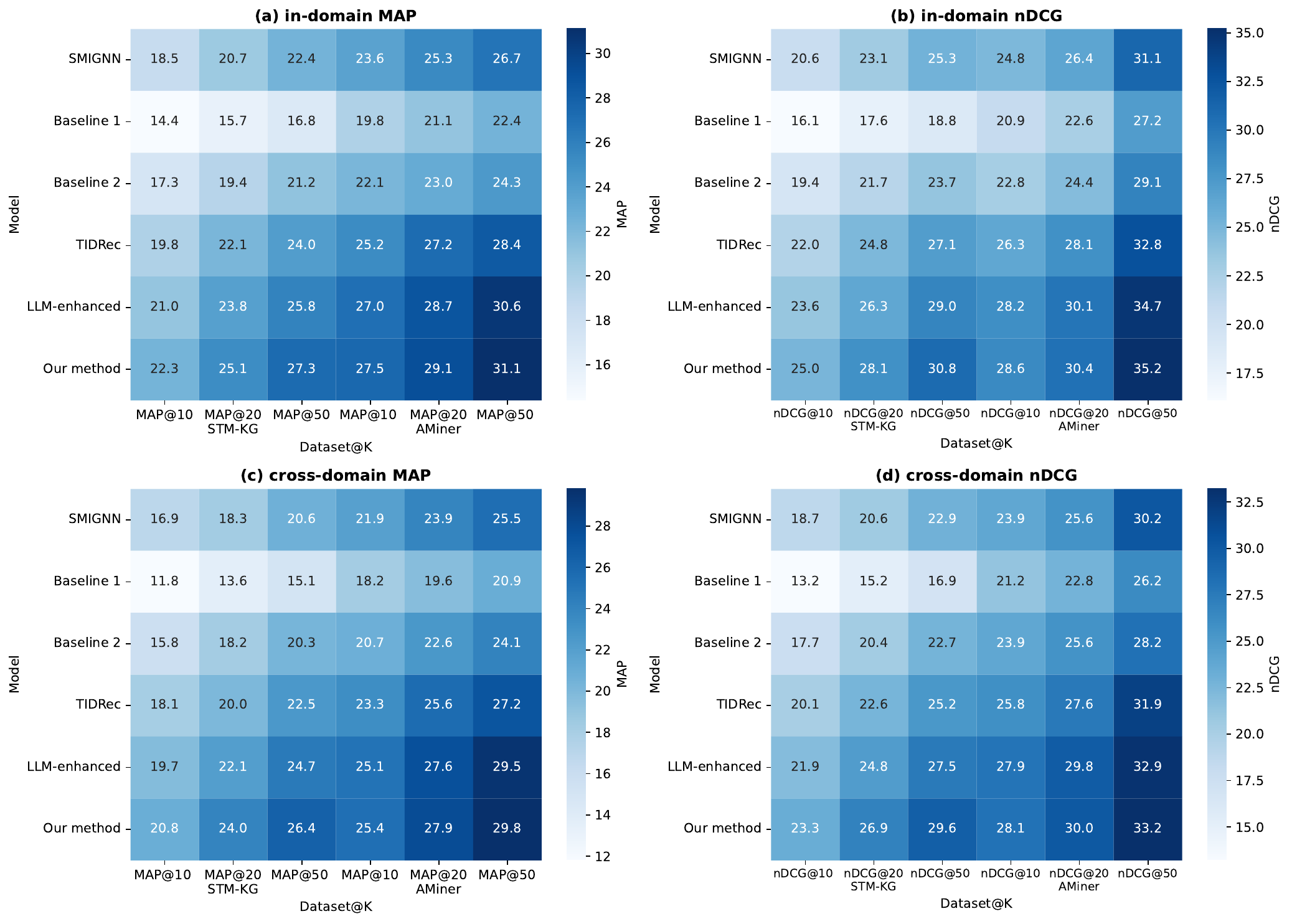}
\caption{\centering Performance on baseline models and our method.}\label{fig6}
\end{figure}

(1) Performance Comparison across Recent Paper Recommendation Methods
Table~\ref{tab:recom_result_baseline} and \Cref{fig6} show the cross-domain and in-domain paper recommendation performance of our method and baseline methods, including results for two evaluation metrics (MAP and NDCG) and two datasets (STM-KG and AMiner 12). For both evaluation metrics, recommendation lists with different $top_k$ values (i.e., $top_{10}$, $top_{20}$, and $top_{50}$) were used to explore the performance of these models. Our method shows significant and statistically significant advantages over the baseline methods on all datasets. Analysis of these tables and charts yields the following insights and analysis:

From the overall recommendation performance, first, our proposed method achieved the highest scores on most of the top ranked correlation indicators, especially in \textit{MAP@50} and \textit{nDCG@50}. This indicates that it has an advantage in retrieving a wider range of semantically related papers.
This performance can be attributed to our model's hybrid representation strategy, which combines:
\begin{itemize} 
\item \textit{GPT-3.5} cross domain embedding is used to capture interdisciplinary contextual language knowledge;
\item Fine grained knowledge entity vectors are used to clarify coding research tasks, methods, and metrics.
\item Fine-grained entities’ text semantic and roles and heterogeneous relations are integrated throughout both recall and ranking, leveraging structure-invariant patterns that transfer across domains.
These design choices enable the system to maintain semantic coherence, even without the deep semantic between text acquired from the large language pretrain process, comparable to recent LLM based methods.
\end{itemize}

\begin{sidewaystable}[htbp]
\centering
\caption{Performance comparison across baseline on two datasets, and the outcomes of the best performing model are shown in bold.}
\label{tab:recom_result_baseline}
\renewcommand{\arraystretch}{1.2} % 行高
\begin{tabular*}{\textwidth}{@{\extracolsep{\fill}}lcccccccccc}
\toprule
Dataset & Scope & Metric & $Top_k$ & Baseline 1 & Baseline 2 & SMIGNN & TIDRec & LLM-enhanced & Our method \\
\midrule
STM-KG & In-domain & MAP & 10 & 14.4 & 17.3 & 18.5 & 19.8 & 21.0 & 22.3 \\
 & & & 20 & 15.7 & 19.4 & 20.7 & 22.1 & 23.8 & 25.1 \\
 & & & 50 & 16.8 & 21.2 & 22.4 & 24.0 & 25.8 & 27.3 \\
 & & nDCG & 10 & 16.1 & 19.4 & 20.6 & 22.0 & 23.6 & 25.0 \\
 & & & 20 & 17.6 & 21.7 & 23.1 & 24.8 & 26.5 & 28.1 \\
 & & & 50 & 18.8 & 23.7 & 25.3 & 27.1 & 29.0 & 30.8 \\
 & Cross-domain & MAP & 10 & 11.8 & 15.8 & 16.9 & 18.1 & 19.7 & 20.8 \\
 & & & 20 & 13.6 & 18.2 & 18.3 & 20.0 & 22.1 & 24.0 \\
 & & & 50 & 15.1 & 20.3 & 20.6 & 22.5 & 24.7 & 26.4 \\
 & & nDCG & 10 & 13.2 & 17.7 & 18.7 & 20.1 & 21.9 & 23.3 \\
 & & & 20 & 15.2 & 20.4 & 20.6 & 22.6 & 24.8 & 26.9 \\
 & & & 50 & 16.9 & 22.7 & 22.9 & 25.2 & 27.5 & 29.6 \\
Aminer v12 & In-domain & MAP & 10 & 19.8 & 22.1 & 23.6 & 25.2 & 27.0 & 27.5 \\
 & & & 20 & 21.1 & 23.0 & 25.3 & 27.2 & 28.7 & 29.1 \\
 & & & 50 & 22.4 & 24.3 & 26.7 & 28.4 & 30.6 & 31.1 \\
 & & nDCG & 10 & 20.9 & 22.8 & 24.8 & 26.3 & 28.2 & 28.6 \\
 & & & 20 & 22.6 & 24.4 & 26.4 & 28.1 & 30.1 & 30.4 \\
 & & & 50 & 27.2 & 29.1 & 31.1 & 32.8 & 34.7 & 35.2 \\
 & Cross-domain & MAP & 10 & 18.2 & 20.7 & 21.9 & 23.3 & 25.1 & 25.4 \\
 & & & 20 & 19.6 & 22.6 & 23.9 & 25.6 & 27.6 & 27.9 \\
 & & & 50 & 20.9 & 24.1 & 25.5 & 27.2 & 29.5 & 29.8 \\
 & & nDCG & 10 & 21.2 & 23.9 & 23.9 & 25.8 & 27.9 & 28.1 \\
 & & & 20 & 22.8 & 25.6 & 25.6 & 27.6 & 29.8 & 30.0 \\
 & & & 50 & 26.2 & 28.2 & 30.2 & 31.9 & 32.9 & 33.2 \\
\botrule
\end{tabular*}
\end{sidewaystable}

Second, LLM-enhanced method ranks second in all four-evaluation metrics and consistently outperforms the results of SMIGNN and TIDRec. This is thanks to its LLM enhanced recommendation strategy. It demonstrates more stronger recommendation performance on Aminer dataset than previous experiment on STM-KG dataset, especially in MAP@50 (30.6) and nDCG@50 (34.7). However, when used directly as a recommender or only for light augmentation, re-rankers primarily optimize text-level semantic similarity on a fixed candidate pool, which is susceptible to long-tail terminology and aliasing prevalent in paper dataset and cannot exploit graph constraints effectively. Consequently, LLM often underperforms strong graph-based methods or hybrid graph-text semantic methods (such as Our method). Some recent study reveals that incorporating structural interaction knowledge can approach or even surpass these performances \citep{li2025chatcrs}. So, it is maybe a good direction for further study in paper recommendation.

Third, as a general-purpose GNN, SMIGNN can capture multiple intents and high-order neighborhoods, generally outperforming baselines that rely solely on textual semantics (such as Baseline 1) or fail to fully utilize graph structure (such as Baseline 2). However, in academic paper recommendation, it lacks explicit relational distillation and domain semantic modeling, such as collaborations between authors and citation relationships between papers. Therefore, it lags behind the three-graph distillation model TIDRec, designed specifically for academic networks, and LLM semantic enhancement.

TIDRec explicitly integrates multiple relational graphs and performs distillation correction, effectively aligning paper content, citation structure, and scholar interests in the paper context. It significantly outperforms general-purpose GNNs (such as SMIGNN) in in-domain paper recommendation. However, since its advantage primarily relies on the consistency of relational distribution within the training domain, its advantage diminishes when crossing domains. Therefore, while this model outperforms SMIGNN in cross-domain paper recommendation, the gap with LLM-enhanced and our method gradually widens.

Baseline 2 enhances the semantic representation of paper content by leveraging inter-paper citation information and fine-grained knowledge semantics. This improves the accuracy of paper recommendation compared to Baseline 1, which relies solely on semantic representation of knowledge sentences.

From the interdisciplinary paper recommendation performance, first, we found that all six methods’ performance degradation in the recommendation lists with same $top_k$ values when applied to cross-domain recommendations. The degree of performance drop varies across models and tends to decrease as the number of recommended papers increases. This trend is evident across MAP and nDCG scores, particularly for smaller K values. For example, nDCG@20 for Baseline 1 decreases from 17.6 (in-domain) to 15.2 (cross-domain), whereas our method decreases from 28.1 to 26.9, indicating a better stable ranking quality. This tendency indicates that in-domain suggestions are generally more accurate than cross-domain recommendations, which could be attributed to semantic differences in sentences, scientific concepts, and entity words between different disciplines. However, when the top-K recommendation list expands, the increased coverage helps offset these semantic differences, thereby partially restoring performance.

Second, among the six methods, Baseline 1 exhibits the largest performance drop, while our proposed method shows the smallest degradation in cross-domain scenarios. This indicates that semantic embeddings of different document information types are unequally affected by disciplinary variation. Although both Baseline 2 and our method rely on structured concept or entity embeddings—which may vary across fields—the incorporation of citation information mitigates the negative effects of semantic divergence. As a result, their performance remains more stable compared to Baseline 1, which does not utilize such structural signals.

(2) In-domain Paper Recommendation Performance with Different Word Embedding Models

\begin{figure}[h]%
\centering
\includegraphics[width=0.9\textwidth]{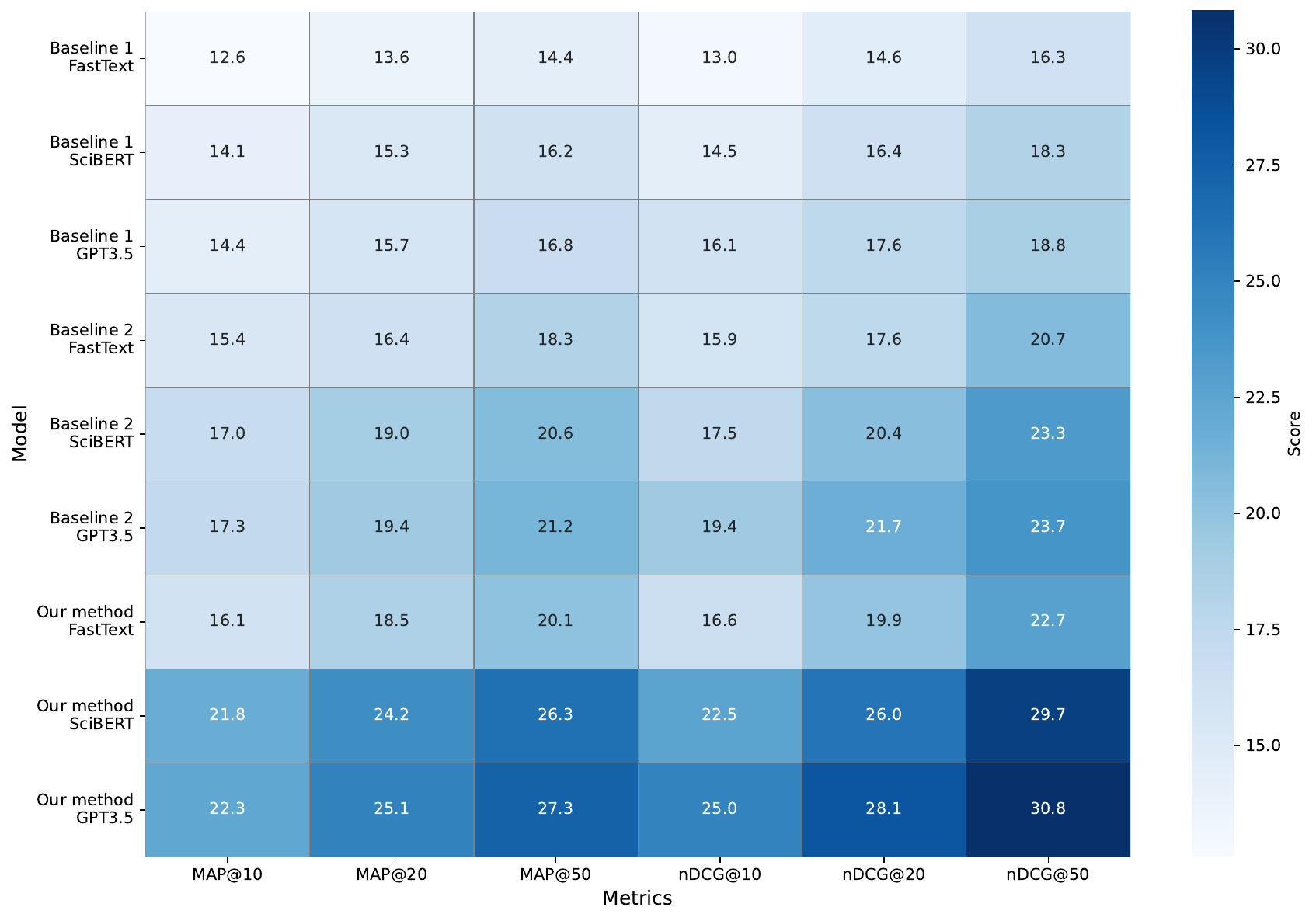}
\caption{\centering Comparison of in-domain paper recommendation performance of our method with
different word embedding models on STM-KG.}\label{fig7}
\end{figure}

Table~\ref{tab:diff_embed_recomm} and \Cref{fig7} show the in-domain paper recommendation performance of our method, Baseline 1 and Baseline 2 which are all mainly depend on text semantic embedding, with different word embedding models, including results for two evaluation metrics (MAP and NDCG) on STM-KG dataset. For both evaluation metrics, recommendation lists with different $top_k$ values (i.e., $top_{10}$, $top_{20}$, and $top_{50}$) were used to explore the performance of these models. Different word embedding models has a significant impact on the performance of our method. After using the pre-trained model embedding of domain-specific vocabulary, the impact of training corpus size on performance is smaller than that of citation information. Analysis of these tables and charts yields the following insights and analysis:

First, in terms of accuracy, our method is sensitive to the choice of embedding model, with \textit{GPT-3.5} consistently achieving the best performance, highlighting the strength of large-scale language models. While upgrading from \textit{FastText} to \textit{SciBERT} results in a significant improvement, the increase from \textit{SciBERT} to \textit{GPT-3.5} is quite minimal. This shows that two variables lead to higher embedding quality: domain-specific vocabulary and training corpus size. Pre-trained language models like \textit{SciBERT} and \textit{GPT-3.5} capture semantic similarity more precisely, resulting in superior recommendation results when employed in embedding-based approaches.

\begin{table}[h]
\centering
\caption{Comparison of performance with different word embedding models, and the outcomes of the best performing model are shown in bold.}
\label{tab:diff_embed_recomm}
\renewcommand{\arraystretch}{1.2} % 调整行高
\begin{tabular}{lcccccc}
\toprule
\multirow{2}{*}{Metric ($Top_k$)} & \multicolumn{3}{c}{MAP} & \multicolumn{3}{c}{nDCG} \\
\cmidrule(lr){2-4} \cmidrule(lr){5-7}
 & 10 & 20 & 50 & 10 & 20 & 50 \\
\midrule
Baseline 1 with FastText & 12.6 & 13.6 & 14.4 & 13.0 & 14.6 & 16.3 \\
Baseline 1 with SciBERT & 14.1 & 15.3 & 16.2 & 14.5 & 16.4 & 18.3 \\
Baseline 1 with GPT3.5 & 14.4 & 15.7 & 16.8 & 16.1 & 17.6 & 18.8 \\
Baseline 2 with FastText & 15.4 & 16.4 & 18.3 & 15.9 & 17.6 & 20.7 \\
Baseline 2 with SciBERT & 17.0 & 19.0 & 20.6 & 17.5 & 20.4 & 23.3 \\
Baseline 2 with GPT3.5 & 17.3 & 19.4 & 21.2 & 19.4 & 21.7 & 23.7 \\
Our method with FastText & 16.1 & 18.5 & 20.1 & 16.6 & 19.9 & 22.7 \\
Our method with SciBERT & 21.8 & 24.2 & 26.3 & 22.5 & 26.0 & 29.7 \\
Our method with GPT3.5 & 22.3 & 25.1 & 27.3 & 25.0 & 28.1 & 30.8 \\
\botrule
\end{tabular}
\end{table}

Second, when using \textit{FastText}, the performance of our method—based on fine-grained knowledge entity embeddings—is lower than that of Baseline 2 (scientific concept embeddings with \textit{SciBERT}). Interestingly, Baseline 2 using \textit{FastText} also outperforms Baseline 1, even when Baseline 1 employs \textit{SciBERT} or \textit{GPT-3.5}. These results suggest that, in the absence of citation information, the choice of embedding model has a greater impact on recommendation performance than the type of document content being embedded. Once citation information is incorporated, however, the relative impact of embedding model selection is reduced, as citation structure provides complementary inter-document association signals that enhance semantic similarity matching.

Third, in terms of ranking, \textit{GPT-3.5} also consistently leads in $nDCG$, with scores of $nDCG@10$ = 25.0, $nDCG@20$ = 28.1, and $nDCG@50$ = 30.8, which are 1.1–2.5 points higher than those achieved by SciBERT. The superior performance of GPT-3.5 is due to its ability to represent complex semantic relationships learned from massive-scale cross-domain corpora, resulting in better top-ranked paper matching. These results confirm that the quality of semantic embedding models significantly affects both retrieval accuracy and ranking effectiveness.

(3) In-domain Paper Recommendation Performance with Different Dimensions of Document Information Embedding

\begin{figure}[h]%
\centering
\includegraphics[width=0.9\textwidth]{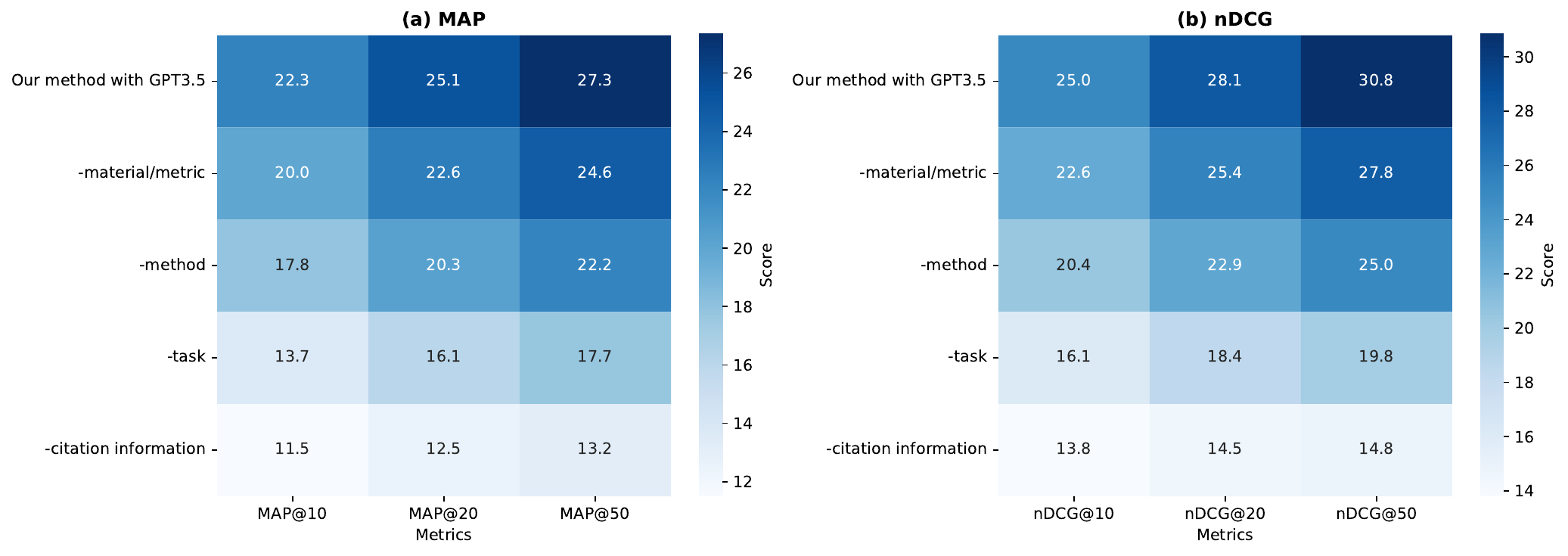}
\caption{\centering Comparison of in-domain paper recommendation performance
Of our method with different dimensions of document information embedding.}\label{fig8}
\end{figure}

\Cref{fig8} show the in-domain paper recommendation performance of our method using different types of document information embeddings, including results for two evaluation metrics (MAP and NDCG) on STM-KG dataset. For both evaluation metrics, recommendation lists with different $top_k$ values (i.e., $top_{10}$, $top_{20}$, and $top_{50}$) were used to explore the performance of these models. We can find out that different types of knowledge entities, and citation information have different degrees of impact on the recommendation performance of our method. Analysis of these tables and charts yields the following insights and analysis:

First, when the semantic embeddings of the fine-grained knowledge entities extracted from paper titles and abstracts (i.e., task, methods, and materials/metrics entities) are successively removed from our method, recommendation performance gradually decreases. When examining each entity type individually, their contribution to recommendation performance follows the order: $\textit{Task} > \textit{Methods} > \textit{Materials}/\textit{Metrics}$. However, the overall reduction is modest compared to the performance of recommendations using semantic embeddings of titles and abstracts alone. This suggests that, compared to titles and abstracts, isolated knowledge entity embeddings do not provide sufficient context for effective recommendations, and that researchers prioritize papers relevant to their research tasks, while focusing on papers with a relatively low relevance for materials and metrics.

Secondly, consider for Table~\ref{tab:recom_result_baseline}, combining the title, abstract, and fine-grained knowledge entities into a semantic embedding improves recommendation accuracy. This is superior to embeddings based solely on the title and abstract, as well as embeddings that use the background and objective statements in the abstract (i.e., Baseline 1). While full abstract embeddings alone offer relatively strong performance due to their rich context, embedding the more targeted and semantically dense research objective statements further enhances the relevance of recommendations. Notably, the inclusion of entity embeddings results in a more fine-grained representation of research intent and methodology, leading to better semantic alignment between papers.

Third, incorporating citation information into document embeddings improves recommendation performance of our method. Consider for Table~\ref{tab:recom_result_baseline} again, Specifically, the SPECTER-based model—which integrates abstracts and citation graphs—performs better than Baseline 1. Building upon this, Baseline 2 further improves performance by combining \textit{SPECTER} embeddings with scientific concept vectors. Our proposed method, which integrates \textit{SPECTER} with structured fine-grained knowledge entities, achieves the highest performance, with a 5.0–6.1-point improvement over Baseline 2. This demonstrates that citation-based structure complements semantic similarity by capturing latent topic relationships, while the coverage and categorization of entities affect embedding quality. Unlike Baseline 2, which lacks coverage of context, purpose, and task-related entities and overlaps between process and method types, our categorization includes key research dimensions—task (what is studied), method (how it is studied), material, and metric (tools and evaluation)—offering a more comprehensive representation of the paper’s core content.

We also assess the ranking quality using nDCG@10, nDCG@20, and nDCG@50, as shown in \Cref{fig8}. The results demonstrate that the abstract + citation + knowledge entity embedding configuration has the highest scores in all nDCG metrics: nDCG@10 = 25.0, nDCG@20 = 28.1, and nDCG@50 =30.8, outperforming Baseline 2 (abstract + citation) by 5.6, 6.4, and 7.1 points, respectively. This demonstrates that the proposed methodology not only boosts overall relevance, but also elevates the most relevant papers to the top, which is more compatible with user expectations in academic search and citation behavior.

To further disentangle the contribution of fine-grained content dimensions, we evaluate three entity-type settings while keeping the encoder fixed: Task-only (\textit{T}), Task+Method (\textit{T+M}), and Task+Method+Material/Metric (\textit{T+M+D}). As summarized in Table~\ref{tab:unified_ablation}, performance improves monotonically from \textit{T → T+M → T+M+D} on \textit{MAP@K} and \textit{nDCG@K} ($K \in [10, 20, 50]$). This indicates that the richer the entity dimension, the more monotonically \textit{MAP} and \textit{nDCG} increase across all \textit{K} values, indicating that semantics related to processes and resources, such as methods/materials, have complementary benefits for task similarity. These results support our design goal of modeling multi-facet semantics that are aligned with scholars’ in-domain information needs. Task entities anchor topical intent; Method and Material/Metric introduce procedural and resource-level distinctions that are frequently decisive for paper similarity in scholarly search. Their inclusion yields more discriminative representations and improves top-\textit{K} ranking quality.

\begin{table*}[t]
\centering
\caption{Unified ablation on entity-type configuration (Section 4.4.2(3)), STM-KG (in-domain). All runs share the same candidate pool and evaluation. Best per column in \textbf{bold}.}
\label{tab:unified_ablation}
\small
\renewcommand{\arraystretch}{1.12}
\begin{tabular}{l l ccc ccc}
\toprule
\multirow{2}{*}{\textbf{Entity Types}} & \multirow{2}{*}{\textbf{Vector Operation}} & \multicolumn{3}{c}{\textbf{MAP@K} $\uparrow$} & \multicolumn{3}{c}{\textbf{nDCG@K} $\uparrow$} \\
\cmidrule(lr){3-5} \cmidrule(lr){6-8}
& & @10 & @20 & @50 & @10 & @20 & @50 \\
\midrule
Title+Abs+Citation       & single vector(doc $s_p$)                & 13.7 & 16.1 & 17.7 & 16.1 & 18.4 & 19.8 \\
\midrule
T                             & single vector $p_t$                          & 17.8 & 20.3 & 22.2 & 20.4 & 22.9 & 25.0 \\
T+M                           & concat $[p_t,p_m]$    & 20.0 & 22.6 & 24.6 & 22.6 & 25.4 & 27.8 \\
T+M+D                       & $p_g=[c_t,c_m,c_d,s_p]$                      & 22.3 & 25.1 & \textbf{27.3} & 25.0 & 28.1 & \textbf{30.8} \\
\bottomrule
\end{tabular}

\vspace{2pt}
\footnotesize
Notes: $p_t=[c_t,s_p]$, $p_m=[c_m,s_p]$, $p_g=[c_t,c_m,c_d,s_p]$.
\end{table*}

Combining the above-mentioned ablation study, we found that the performance of semantic vector similarity-based paper recommendation is influenced by three main factors:
\begin{itemize}
\item The extent to which the embedded content reflects the core research dimensions of the paper, and 
\item The capability of the semantic embedding model, and
\item The inclusion of citation information.
\end{itemize}

Among these, the total contribution of each type of fine-grained entities has the greatest impact on the performance improvement of paper recommendation, followed by the choice of embedding model, and finally the citation information embeddings.

(4) Performance of Paper Recommendation Models with Vector Similarity Combination and Weights

\begin{figure}[h]%
\centering
\includegraphics[width=0.9\textwidth]{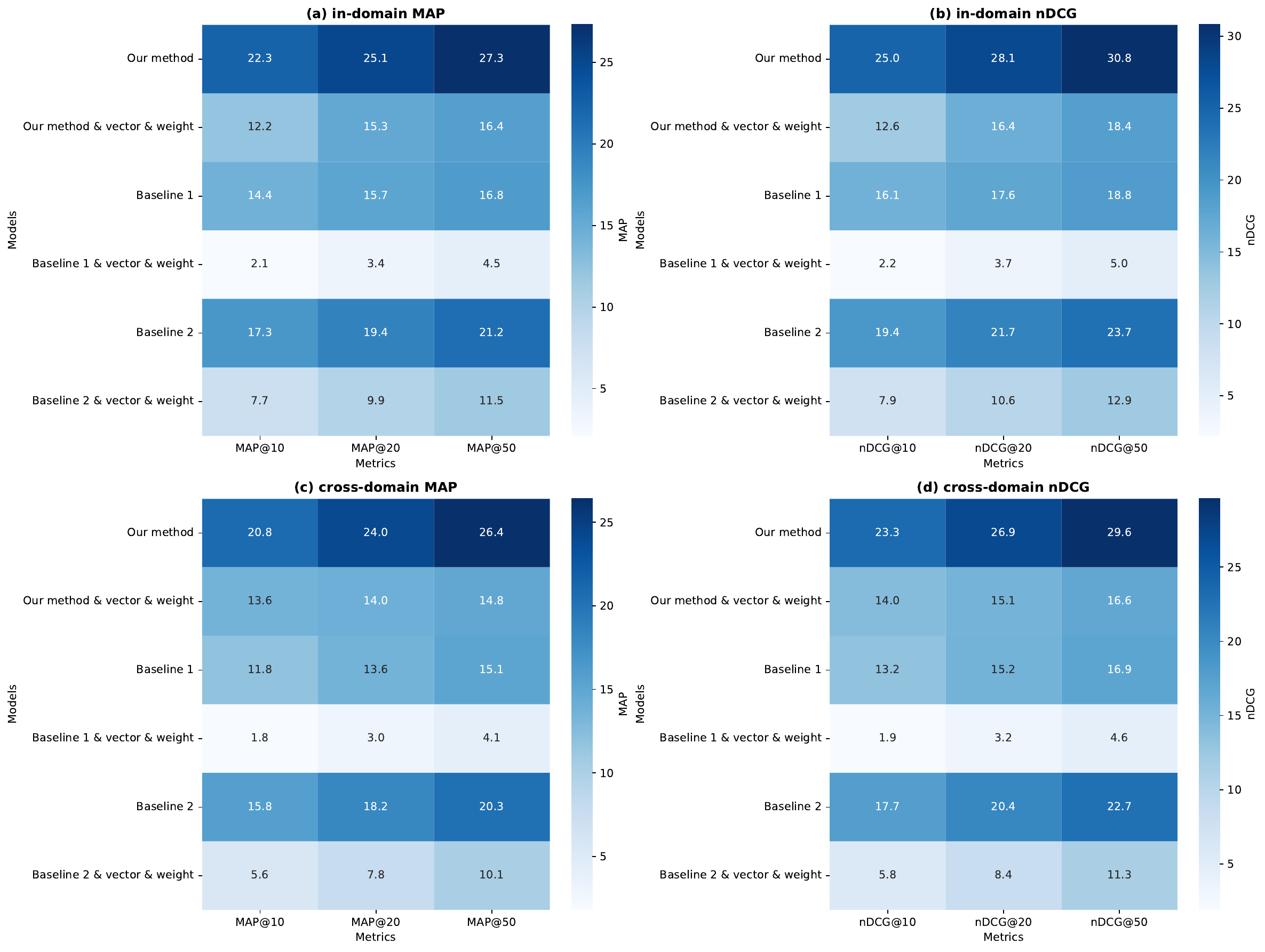}
\caption{\centering Performance comparison of paper recommendation models
with vector similarity combination and weights.}\label{fig9}
\end{figure}

\Cref{fig9} show the cross-domain and in-domain paper recommendation performance of our method, Baseline 1 and Baseline 2 which are all mainly depend on text semantic embedding, incorporating vector similarity combinations and weighting strategies, including results for two evaluation metrics (MAP and NDCG). For both evaluation metrics, recommendation lists with different $top_k$ values (i.e., $top_{10}$, $top_{20}$, and $top_{50}$) were used to explore the performance of these models. After our method is incorporated with similarity-weighted vector combinations, the recommendation performance evaluation indicators are all reduced, and the reduction rate are higher in the in-domain recommendation. Analysis of these tables and charts yields the following insights and analysis:

First, the MAP@N scores of all methods drop considerably after applying similarity-weighted vector combinations and $nDCG@20$ of our method (cross-domain) drops from 26.9 to 15.1. Despite this decline, our method still outperforms baselines, maintaining nDCG@50= 16.6, higher than other methods. This indicates a reduced overlap between the recommended paper list and the reference list of the query paper. From another perspective, this implies that the recommendations begin to deviate from known citations, potentially providing unique insights into research tasks, methodologies, or solutions. 
The basic logic is that when a certain entity type (e.g., task) is highly similar but other entity types are dissimilar, weighting reduces the overall similarity value. As a result, top-ranked recommended papers have asymmetric relevance, stressing one component (e.g., task) but deviating in others—introducing diversity and unique viewpoints on the same research task.

Second, our proposed method consistently achieves the highest MAP@N scores under the weighted setting and exhibits a smaller performance drop compared to Baseline 1 and Baseline 2. This suggests that our method aligns more closely with scholars’ actual citation behavior, favoring papers that share high semantic similarity across multiple dimensions. However, this also implies that our method is relatively less effective in recommending diverse papers, as it tends to favor similarity over divergence.

This observation raises an important research question: should recommendation performance prioritize accuracy—as measured by citation overlap—or diversity, reflected by differences from cited works?
Additionally, the trends observed here are consistent with the results from previous experiments:
\begin{itemize}
\item In-domain recommendation performance remains higher than cross-domain;
\item $MAP@N$ scores increase with a larger number of recommended papers.
\end{itemize}

We evaluate the vector operations exactly as defined in \S3.4 without introducing external 
aggregation schemes. Concretely, we compare:
\begin{itemize}
    \item Single-vector: using only $p_g = [c^t_p, c^m_p, c^d_p, s_p]$ (All-entities) by concatenating. 
    \item Unweighted convex combination: equal weights over $(p_g, p_t, p_m, p_d)$ in Eq.~(7)
    \item Accuracy-tuned weights: learning $w$ in Eq.~(7) on the dev split (MAP@50 objective, 
          $\lambda = 0$) then fixing at test time.
    \item Task-subset re-ranking: within $CP_{\text{task}}$, learning $\boldsymbol{\alpha}$ in Eq.~(9) to 
          aggregate $\{s_j \}_{j=1}^4$  (same/different Method; same/different Material/Metric), applied on top of 
          $\mathbf{w}$.
\end{itemize}

Results in Table~\ref{tab:combine_ablation} show that on STM-KG (in-domain), the single-vector representation employing $p_g$ achieves optimal \textit{MAP}/\textit{nDCG} performance. As the aggregation strategy transitions from uniform weighting ($w=\frac{1}{4}\mathbf{1}$), learned weights (Eq.~(7), $\lambda=0$) to learned weights + $\boldsymbol{\alpha}$ re-ranking (Eq.~(9)), \textit{ILD} and \textit{Coverage} exhibit consistent improvement while accuracy declines correspondingly---this typifies the canonical accuracy--diversity trade-off. Overall, $p_g$ is well-suited for precision-oriented retrieval scenarios; uniform weighting and re-ranking configurations deliver enhanced result diversity and coverage at larger \textit{K} values (e.g., @50), thereby proving more beneficial for exploratory retrieval contexts, including method expansion and cross-perspective comparison.

\begin{table}[t]
\centering
\caption{Vector similarity combination / weighting strategies defined in \S3.4 (Section 4.4.2(4)), STM-KG (in-domain). All runs share the same candidate pool and evaluation. Best per column in \textbf{bold}.}
\label{tab:combine_ablation}
\small
\setlength{\tabcolsep}{3.2pt} % reduce column separation
\renewcommand{\arraystretch}{1.12}
\begin{tabular}{l l ccc ccc}
\toprule
\multirow{2}{*}{\textbf{Combination Types}} & \multirow{2}{*}{\textbf{Vector Operation}} & \multicolumn{3}{c}{\textbf{MAP/ILD@K}} & \multicolumn{3}{c}{\textbf{nDCG/Coverage@K}} \\
\cmidrule(lr){3-5} \cmidrule(lr){6-8}
& & @10 & @20 & @50 & @10 & @20 & @50 \\
\midrule
T+M+D       & $p_g=[c^t_p,c^m_p,c^d_p,s_p]$                & 22.3$/$28.9 & 25.1$/$30.7 & 27.3$/$32.4 & 25.0$/$29.8 & 28.1$/$33.5 & 30.8$/$37.2 \\
T+M+D                       & unweighted: $w=\tfrac{1}{4}\mathbf{1}$   & 19.3$/$30.4 & 21.6$/$32.1 & 24.1$/$35.6 & 21.3$/$31.2 & 24.0$/$26.0 & 26.6$/$40.2 \\
T+M+D                       & \textit{learned $w$} ($\lambda{=}0$)   & 15.2$/$37.1 & 16.9$/$38.4 & 18.7$/$39.6 & 16.3$/$37.9 & 18.3$/$41.0 & 20.1$/$44.1 \\
T+M+D, $CP_{task}$          & \textit{learned $w$} + \textit{$\boldsymbol{\alpha}$} & 12.2$/$\textbf{38.6} & 15.3$/$\textbf{43.1} & 16.4$/$\textbf{47.2} & 12.6$/$\textbf{40.4} & 16.4$/$\textbf{45.7} & 18.4$/$\textbf{50.1} \\
\bottomrule
\end{tabular}

\vspace{2pt}
\footnotesize
Notes: (1) Eq.~(7) forms $p=\sum_{i\in\{g,t,m,d\}} w_i p_i$ with $w_i\!\ge\!0,~\sum w_i\!=\!1$; (2) Eq.~(9) aggregates $\{s_j\}_{j=1}^4$ inside $CP_{task}$ with $\alpha_j\!\ge\!0,~\sum \alpha_j\!=\!1$.
\end{table}

(5) Complementarity with LLM based recommendation systems

The most recent breakthroughs in Large Language Models (\textit{LLMs}), such as \textit{GPT-3.5} and \textit{GPT-4}, enable remarkable capabilities for creating, retrieving, and recommending academic content based on unstructured inputs such as paper titles, abstracts, or searches. Although LLMs perform well in general recommendation or question-answering tasks, their use in academic paper recommendation presents issues in terms of transparency, authenticity, and control.
Our proposed method complements LLM in the following aspects:
\begin{itemize}
\item Structured semantic representation: Our approach does not rely solely on end-to-end text generation, but employs fine-grained scientific entity extraction (tasks, methods, metrics) and constructs vector representations that preserve semantic structure. This makes controllable recommendation strategies (such as "same task, new method") and interpretable retrieval possible.

\item Interpretability and Customization: Entity based vector concatenation allow users to trace recommendation results back to specific semantic components (e.g. task similarity, method differences). In contrast, LLM typically operates like a black box and does not support this fine-grained customization unless a large amount of rapid engineering design is done.

\item Efficiency and deploy ability: Our method requires less computation, no API access or quick tuning, and supports batch computing and offline indexing, making it more suitable for large-scale or institutional deployment.
\end{itemize}

Nevertheless, we believe that LLM can enhance our processes in future work. For example:

\begin{itemize}
\item LLM can assist in more accurate entity disambiguation and relation type induction;

\item LLM can generate user intent embeddings or assist in zero sample query extensions;

\item The feedback generated by LLM can be used to fine tune weight strategies in multi-faceted vector concatenation.
\end{itemize}

In summary, our approach is a structured and interpretable foundation that benefits from integration with LLM functionality, rather than a competitive alternative. In future work, we will investigate a hybrid recommendation framework that integrates knowledge embedding representation and generative techniques.

(6) Text-only comparisons under STM-KG and in-domain

Table~\ref{tab:t_test_recomm} show that $\textit{BM25 (text-only)}< \textit{SciBERT (text-only)} < \textit{Our (text-only encoder)} < \textit{Our (full)}$. \textit{Our (full)} significantly outperforms the strong text baseline \textit{SciBERT} in both $MAP@50$/$nDCG@50$ and entity-awareness correlation ($p < 0.01$ for all three metrics), indicating that relying solely on textual representations is insufficient for optimal performance. Retaining our pipeline and replacing it with a pure text encoder still outperforms \textit{SciBERT}, demonstrating that the fine-grained entity signal and heterogeneous structure inherently contribute to gains. Further ablation analysis shows that removing citation-side structural features results in a slight drop in performance from \textit{Our (full)}, but still outperforms \textit{SciBERT} and \textit{BM25}, and the ranking of entity-awareness metrics is consistent with the ranking metrics. These results demonstrate that the advantage of our approach stems not from potential citation leakage but rather from the synergistic effect of entity, structure, and text; citation information contributes positively but not exclusively.

\begin{table}[h]
\centering
\caption{Text-only comparisons under the temporal decontamination protocol (STM-KG, in-domain). Mean$\pm$std over 3 runs; paired $t$-test vs. SciBERT.}
\label{tab:t_test_recomm}
\renewcommand{\arraystretch}{1.2} % 调整行高
\begin{tabular}{lccc}
\toprule
Setting & MAP@50 & nDCG@50 & Entity-aware Rel. \\
\midrule
BM25 (text-only) & $22.6{\small\pm0.1}$ & $23.0{\small\pm0.1}$ & $0.412{\small\pm0.006}$ \\
SciBERT (text-only) & $24.7{\small\pm0.1}$ & $25.3{\small\pm0.1}$ & $0.441{\small\pm0.005}$ \\
Our (text-only) & $26.3{\small\pm0.1}$ & $29.7{\small\pm0.1}$ & $0.489{\small\pm0.005}$ \\
\textbf{Our (full)} & $\mathbf{27.3}{\small\pm0.1}$ & $\mathbf{30.6}{\small\pm0.1}$ & $\mathbf{0.507}{\small\pm0.004}$ \\
\quad \emph{$p$-value vs. SciBERT} & $<\!0.01$ & $<\!0.01$ & $<\!0.01$ \\
\midrule
\multicolumn{4}{l}{\emph{Ablation: remove citation-side structure (no co-citation / bib.-coupling)}} \\
Our (full) $\rightarrow$ no citation-graph & $26.6{\small\pm0.1}$ & $29.9{\small\pm0.1}$ & $0.498{\small\pm0.004}$ \\
\bottomrule
\end{tabular}
\end{table}

(7) Stratified and balanced evaluations on popularity bias

Table~\ref{tab:stra_balan_recomm} show that absolute scores increase with citation count—as expected—but relative improvements of Our (full) vs. \textit{SciBERT} persist across all strata and are smallest in the highest-cited tertile, indicating a modest yet non-deterministic popularity effect. Results on the balanced subset mirror the stratified analysis, suggesting that our conclusions do not hinge on highly-cited items.

\begin{table}[h]
\centering
\caption{Stratified and balanced evaluations to assess popularity bias (STM-KG, MAP@50). Mean$\pm$std over 3 runs.}
\label{tab:stra_balan_recomm}
\renewcommand{\arraystretch}{1.2} % 调整行高
\begin{tabular}{lccc}
\toprule
Subset & SciBERT (text-only) & Our (full) & $\Delta$ \\
\midrule
Low-cited tertile & $23.6{\small\pm0.1}$ & $\mathbf{26.2}{\small\pm0.1}$ & $+2.6$ \\
Med-cited tertile & $24.9{\small\pm0.1}$ & $\mathbf{27.4}{\small\pm0.1}$ & $+2.5$ \\
High-cited tertile & $25.6{\small\pm0.1}$ & $\mathbf{28.3}{\small\pm0.1}$ & $+2.7$ \\
\midrule
Balanced subset (top-coded) & $24.5{\small\pm0.1}$ & $\mathbf{27.0}{\small\pm0.1}$ & $+2.5$ \\
\bottomrule
\end{tabular}
\end{table}

\subsubsection{Diversity-aware Evaluation on Academic Paper Recommendation}\label{s3subsubsec3}

We evaluates a diversity-aware setting: starting from accuracy-tuned scores \(\lambda=0\) and a shared candidate pool, we apply a re-ranking objective-- 
$
J_{\text{div}} = \text{\textit{MAP}}@50 + \lambda \cdot \text{\textit{ILD}}@50
$
with \(\lambda > 0\) to explicitly enhance intra-list diversity (\textit{ILD}) and coverage.
We introduced two metrics for recommendation diversity, compared and analyzed the changes in recommendation accuracy and diversity values of Baseline 1, Baseline 2 and the recommendation method proposed in this paper for whether using vector similarity combinations and weights or not, and further discussed the trade-off between recommendation accuracy and diversity and their respective main application scenarios. 
Both metrics are calculated based on the cosine distance between document embeddings. For $ILD$, we represent each paper using the \textit{Task} + \textit{Method} + \textit{Material}/\textit{Metric} vector concatenation.

(1) The Definition of Diversity in Scientific Paper Recommendations

In the context of academic paper recommendation, diversity refers to the degree to which recommended articles differ from each other in the list of papers that accurately match the scholar's query and cover multiple dimensions such as topics, particularly in areas such as methodology, material and metrics. We define diversity operationally as recommending articles that have the same research task as the query paper but differ considerably in methodologies or material/metrics, so providing alternate viewpoints or unique approaches to a well-known topic.

This approach is consistent with scenarios such as early-stage subject inquiry, interdisciplinary literature discovery, or methodological extension, in which the goal is not simply to retrieve what is already known (i.e., high \textit{MAP} and \textit{nDCG} overlap), but to extend intellectual exposure.

(2) Experimental Results and Analysis

We further investigate how the combination of vector similarity and weighting influences both recommendation accuracy and diversity. Under the setting where knowledge entities are embedded using \textit{GPT-3.5}, we compare the performance of Baseline 1, Baseline 2 which mainly use text embedding and combine vector with learned weight, and our proposed method in the in-domain paper recommendation scenario, aiming to reveal patterns in how these factors affect accuracy and diversity.

\begin{figure}[h]%
\centering
\includegraphics[width=0.9\textwidth]{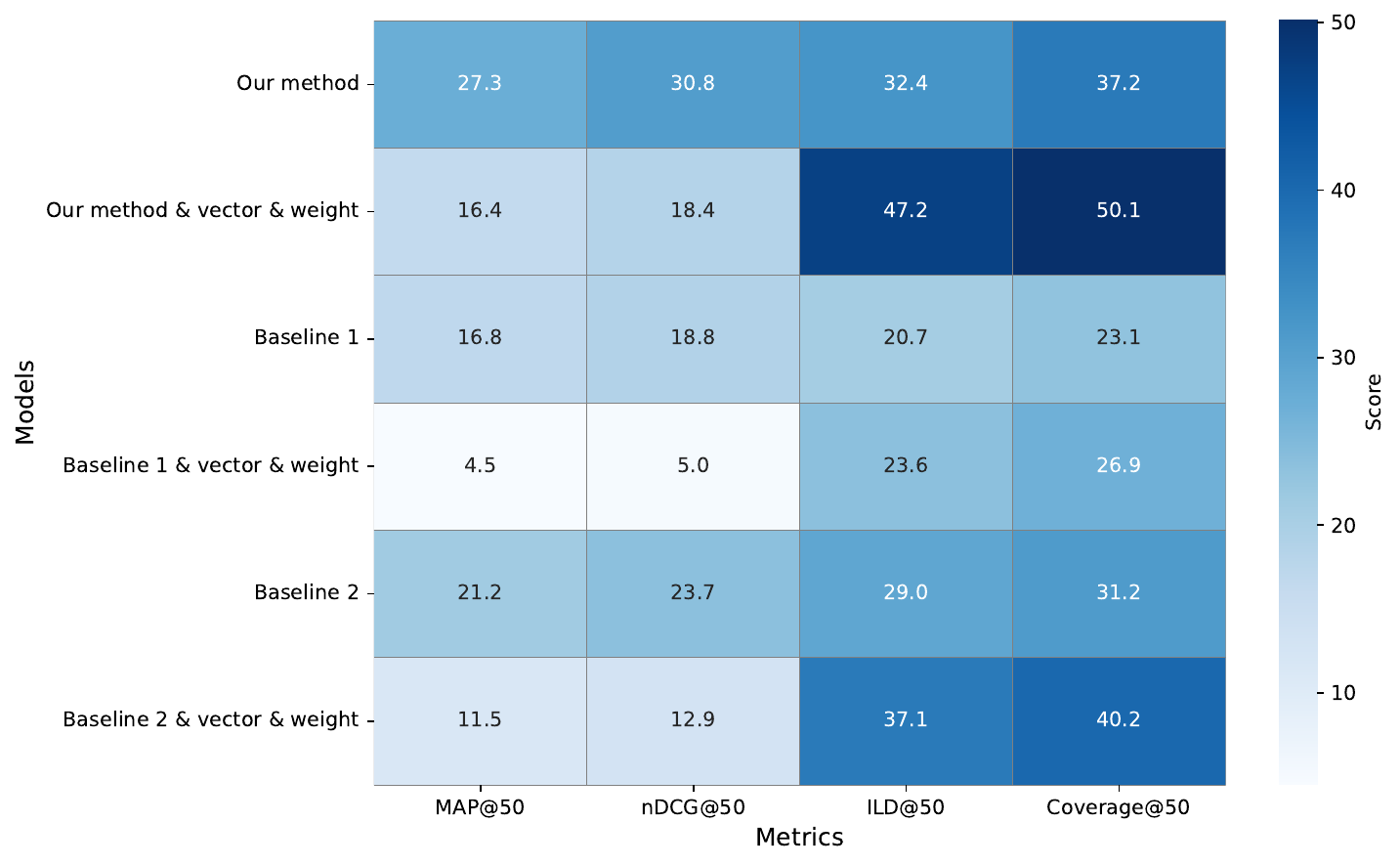}
\caption{ \centering{ Accuracy and Diversity comparison of paper recommendation models with vector similarity
combination and weights, diversity-aware re-ranking ($\lambda > 0$).}}\label{fig10}
\end{figure}

As shown in \Cref{fig10}, applying vector similarity combination and weighting results in a decrease in recommendation accuracy across all three methods—consistent with the observations in performance comparison of paper recommendation models with vector similarity combination and weights. At the same time, we observe improvements in both diversity-related metrics. For our proposed method, the $ILD$ score increases from 32.4 to 47.2, and the $Coverage$ metric rises from 37.2 to 50.1. These improvements are more substantial than those seen in the other two baselines. Compared with single-vector and unweighted multi-vector baselines, the diversity-aware setting substantially improves \textit{ILD} and \textit{Coverage}. The slight decrease in \textit{MAP}/\textit{nDCG} reflects the expected trade-off induced by explicitly optimizing a diversity component.

Taken together with the findings from comparison of in-domain paper recommendation performance of our method with different word embedding models and comparison of the cross-domain and in-domain paper recommendation performance, we attribute this advantage to our method’s superior semantic representation of knowledge entities, enabled by \textit{GPT-3.5} embeddings. This allows the model to better distinguish and match papers with different or similar tasks, methods, and materials/metrics—thus achieving a better balance between accuracy and diversity.

The results demonstrate that our weighted combination technique improves $ILD$ and $Coverage$, resulting in a more diversified suggestion list, albeit at the expense of a minor decrease in accuracy ($MAP$). This exemplifies the typical trade-off between accuracy and diversity.

(3) Application Scenarios: Accuracy versus Diversity

We also underline the importance of aligning recommendation strategies with user scenarios. High $MAP$ and $nDCG$ values are advised for predicting citations or maintaining established lines of work. For concept generation, interdisciplinary research, or early-stage evaluation, diversity-aware rankings may be more beneficial.

This study shows that utilizing diversity-aware vector weighting produces more diverse paper recommendations, which can better meet exploratory research objectives. Future research could include adaptive re-ranking algorithms that weigh accuracy and novelty based on user preferences or task categories.

\subsubsection{Experimental summary on Our method on Academic Paper Recommendation}\label{s3subsubsec4}

In this paper, we validate that integrating multiple types of document information—such as titles, abstracts, citation graphs, and fine-grained knowledge entities—consistently improves the performance of baseline paper recommendation methods.

Incorporating semantic embeddings generated by large language models, particularly \textit{GPT-3.5}, further enhances recommendation accuracy, especially when applied to document components like titles, abstracts, and knowledge entities.

Our analysis shows that while \textit{SPECTER} performs slightly better for in-domain knowledge mapping, knowledge entity embeddings derived from interdisciplinary knowledge graphs contribute more significantly to cross-domain recommendation performance.

Moreover, the proposed \textit{SPECTER} + knowledge entity embeddings + vector weighting method achieves the best \textit{MAP@K} scores under weighted settings—particularly at \textit{MAP@50}, where it consistently outperforms all other vector-weighted baselines.

Under the accuracy-tuned setting ($\lambda=0$), the proposed \textit{SPECTER} + knowledge entity embeddings + vector weighting method achieves the best or statistically non-inferior \textit{MAP@K}/\textit{nDCG@K} against single-vector and unweighted multi-vector baselines. In contrast, \Cref{fig10} reports the diversity-aware re-ranking ($\lambda>0$), where explicit diversity promotion leads to significant \textit{ILD}/Coverage gains with a small, controlled decrease in \textit{MAP}/\textit{nDCG}.

We attribute this improvement to the flexibility of weighted similarity combination, which allows the recommendation system to surface papers that not only match the cited references but also provide novel and diverse perspectives. This balance between accuracy and novelty enables the system to better support scholars’ varying information needs.

In summary, our experiments demonstrate that the integration of fine-grained knowledge graphs as an auxiliary semantic source significantly enhances paper recommendation performance. This enhancement holds across different embedding strategies, whether using full abstract representations or structured embeddings of task, method, and material/metric entities.

\section{Conclusion and Future Works}\label{sec5}
This paper presents a new approach to academic paper recommendation, aiming to address the limitations of existing methods that struggle to align with scholars' specific literature needs and provide sufficiently broad results. The study constructs an FG-SKGs by extracting fine-gained knowledge entities from paper titles and abstracts, capturing their relations and citation links. To achieve this, the authors employ a combination of \textit{SciBERT}, \textit{BiLSTM}, and Cascade models to extract fine-grained entities and map their relations. The Transform model is used to integrate abstract and citation data into meaningful embeddings, while \textit{GPT-3.5} supports the semantic embedding of knowledge entities. The recommendation process relies on multidimensional vector operations, representing papers through knowledge entity relations. By applying vector filtering and ranking techniques, the system recommends papers that align with scholars' specific needs.
The experimental results demonstrate improvements in hit accuracy and diversity, providing more relevant and diverse recommendations. This approach not only aligns with scholars' literature needs but also broadens their research perspectives. Future work includes further refining the relations between entities to capture complex associations and enhancing graph representation learning. The study aims to balance the objective evaluation of recommendations with scholars' subjective input to enhance the accuracy and diversity of the recommendations.

\bmhead{Acknowledgements}
This work is supported by the youth project of the National Social Science Fund of China, titled " Identifying the Semantic Relationship and Evolutionary Paths of Algorithm Entities Based on the Full-text of Academic Papers" (Grant No. 24CTQ027). This paper further expands on the relevant research status, research methods, and experimental discussions based on the JCDL2024 conference paper.

\section*{Declarations}
The author(s) declared no potential conflicts of interest with respect to the research, authorship, and/or publication of this article.

\bibliography{sn-bibliography}% common bib file

\appendix
\section*{Appendix A: Annotation Guideline for Entity Identification}
\addcontentsline{toc}{section}{Appendix A: Annotation Guideline for Entity Identification}

\subsection*{A.1 Overview}
This annotation task aims to identify seven types of fine-grained knowledge entities from the titles and abstracts of scientific papers, including: Task, Object, Problem, Method, Process, Material and Metric.
All entities are annotated at the sentence level using the standard BIO (Begin-Inside-Outside) annotation format. The annotation results will be used to build a fine-grained knowledge graph and support downstream recommendation tasks.

\subsection*{A.2 Entity Category}
\paragraph{1. Task}
\begin{itemize}
  \item \textbf{Definition:} The overall goal or task of the research, usually answering the question of ``what to do''.
  \item \textbf{Typical location:} the middle of title, beginning of abstract or purpose sentence.
  \item \textbf{Keyword clues:} classification, prediction, generation, detection, analysis, modeling, estimation, etc.
  \item \textbf{Example:}
  \begin{itemize}
    \item We propose a new model for \fcolorbox{white}{green}{\textit{machine reading comprehension}}.
    \item This work focuses on \fcolorbox{white}{green}{\textit
    {sentiment analysis}} of product reviews.
  \end{itemize}
  \item \textbf{Labeled as:} B-TASK, I-TASK
\end{itemize}

\paragraph{2. Object}
\begin{itemize}
  \item \textbf{Definition:}The specific object or entity operated, processed or analyzed in the study, limited the scope of the research task, that is, 'who' is studied.
  \item \textbf{Typical location:} immediately after the task word or method word.
  \item \textbf{Keyword clues:} text, image, gene, question, article, protein, sentence, molecule, etc.
  \item \textbf{Example:}
  \begin{itemize}
    \item We propose a new model for machine reading comprehension on \fcolorbox{white}{green}{\textit{biomedical documents}}.
    \item The system classifies \fcolorbox{white}{green}{\textit{customer reviews}} into positive or negative.
  \end{itemize}
  \item \textbf{Labeled as:} B-OBJECT, I-OBJECT
\end{itemize}

\paragraph{3. Problem}
\begin{itemize}
  \item \textbf{Definition:} The core problem, difficulty or pain point to be solved in the study, emphasizing "what difficulties exist".
  \item \textbf{Typical location:} the front of the abstract or the description of the research motivation.
  \item \textbf{Keyword clues:} challenge, limitation, bottleneck, issue, shortcoming, drawback, etc.
  \item \textbf{Example:}
  \begin{itemize}
    \item One key challenge in neural parsing is \fcolorbox{white}{green}{\textit{domain generalization}}.
    \item Existing methods suffer from \fcolorbox{white}{green}{\textit{low robustness to noise}}.
  \end{itemize}
  \item \textbf{Labeled as:} B-PROBLEM, I-PROBLEM
\end{itemize}

\paragraph{4. Method}
\begin{itemize}
  \item \textbf{Definition:} The method, model, algorithm, technology, and architecture used to complete the research task.
  \item \textbf{Typical location:} in the middle or at the end of the sentence, often co-occurring with verbs such as propose/use / apply/develop.
  \item \textbf{Keyword clues:} model, framework, architecture, algorithm, approach, technique, etc.
  \item \textbf{Example:}
  \begin{itemize}
    \item We design a novel \fcolorbox{white}{yellow}{multi-head self-attention mechanism}.
    \item Our approach utilizes a \fcolorbox{white}{yellow}{graph convolutional network}.
  \end{itemize}
  \item \textbf{Labeled as:} B-METHOD, I-METHOD
\end{itemize}

\paragraph{5. Process}
\begin{itemize}
  \item \textbf{Definition:} The behavior of changing or generating state information of the research object in research methods.
  \item \textbf{Typical location:} when introducing the research method or as a model simulation steps.
  \item \textbf{Keyword clues:} photosynthesis, protein folding, diffusion, signal transduction, fermentation, etc.
  \item \textbf{Example:}
  \begin{itemize}
    \item We simulate the \fcolorbox{white}{yellow}{protein–protein interaction process using a GNN model}.
    \item Our study models the \fcolorbox{white}{yellow}{carbon cycle in tropical forests}.
  \end{itemize}
  \item \textbf{Labeled as:} B-PROCESS, I-PROCESS
\end{itemize}

\paragraph{6. Material}
\begin{itemize}
  \item \textbf{Definition:} External resources used for experiments or evaluations, training or data analysis, including datasets, tools, software, and equipment, etc.
  \item \textbf{Typical location:} in the experimental description, at the end of the sentence or in the brackets.
  \item \textbf{Keyword clues:} dataset, corpus, system, simulator, API, software, device, platform, etc.
  \item \textbf{Example:}
  \begin{itemize}
    \item Experiments are conducted on the \fcolorbox{white}{cyan}{CIFAR-10 dataset}.
    \item We implemented our model using \fcolorbox{white}{cyan}{PyTorch}.
  \end{itemize}
  \item \textbf{Labeled as:} B-MATERIAL, I-MATERIAL
\end{itemize}

\paragraph{7. Metric}
\begin{itemize}
  \item \textbf{Definition:} An indicator used to evaluate model performance or compare methods.
  \item \textbf{Typical location:} experimental results description or end.
  \item \textbf{Keyword clues:} $BLEU$, $ROUGE$, $F1$, $accuracy$, $precision$, $recall$, $RMSE$, etc.
  \item \textbf{Example:}
  \begin{itemize}
    \item We evaluate the model using \fcolorbox{white}{magenta}{$BLEU$} and \fcolorbox{white}{magenta}{$ROUGE$} scores.
    \item The best \fcolorbox{white}{magenta}{accuracy} achieved is 94.5\% on the test set.
  \end{itemize}
  \item \textbf{Labeled as:} B-METRIC, I-METRIC
\end{itemize}

\subsection*{A.3 Boundary Determination Rules}

\begin{itemize}
  \item When encountering a composite entity (such as "image classification task"), the entire phrase is annotated; Nested entities: only the outermost entity is annotated, and no nesting is performed:
  \item When a term is interrupted (such as a method name is inserted with a modifier), it can be annotated across short stops, such as: "a novel convolutional neural network based approach" $\rightarrow$
  the whole is marked as METHOD;
  \item The following types are not annotated: author name, year, predicate verb, fuzzy word (such as "approach" if the technology is not specified, it is not annotated):
  \item Adjective pronouns (this, its, these, such) and determinators (the, a) should not be included in the span.
\end{itemize}

\section*{Appendix B: Pseudocode of Our Proposed Paper Recommendation Method.}
\addcontentsline{toc}{section}{Pseudocode of Our Proposed Paper Recommendation Method}

\begin{algorithm}[h]
\caption{Entity-aware Multi-Vector Paper Recommendation}
\begin{algorithmic}[1]
\State \textbf{Inputs:}
\State \quad $q$ -- query paper
\State \quad $P$ -- set of papers
\State \quad $G=(P,E,L_{PP},L_{PE},L_{EE})$ -- FG-SKG: papers, entities, citations, paper-entity links, entity-entity links
\State \quad $K$ -- candidate pool size (top-$K$)
\State \quad $N$ -- number of recommendations to return (top-$N$)
\State

\State // \textbf{A. Node Embeddings \& Entity Aggregation} [\S3.3.1, Eqs.~(1)(3)]
\For{each paper $p \in P$}
    \State $s_p \gets \text{SPECTER}(\text{title/abstract of } p)$
    \State // Encode entity texts with GPT and aggregate by sum pooling
    \State $c_p^t \gets \sum_{e \in \mathcal{T}_p} \text{GPT}(e)$
    \State $c_p^m \gets \sum_{e \in \mathcal{M}_p} \text{GPT}(e)$
    \State $c_p^d \gets \sum_{e \in \mathcal{D}_p} \text{GPT}(e)$
\EndFor
\State

\State // \textbf{B. Multi-Vector Composition} [\S3.3.2, Eqs.~(4)(6)]
\For{each paper $p \in P$}
    \State $p_g \gets [c_p^t, c_p^m, c_p^d, s_p]$
    \State $p_t \gets [c_p^t, s_p]$
    \State $p_m \gets [c_p^m, s_p]$
    \State $p_d \gets [c_p^d, s_p]$
\EndFor
\State

\State // \textbf{C. Candidate Generation (Coarse)} [\S3.4, Eqs.~(7)(8)]
\For{each paper $p \in P$}
    \State $p^* \gets \sum_{i \in \{g,t,m,d\}} w_i \cdot p_i$ \quad // $w_i \ge 0, \sum_i w_i = 1$ (learned on dev set)
    \State $\text{score} \gets \cos(q, p^*) = \dfrac{q^\top p^*}{\|q\| \cdot \|p^*\|}$
\EndFor
\State $CP \gets$ top-$K$ papers by score
\State

\State // \textbf{D. Task-Aware Refinement (Fine)} [\S3.4, Eq.~(9)]
\State $CP_{\text{task}} \gets$ task-similar subset of $CP$
\For{each paper $p \in CP_{\text{task}}$}
    \State $p_{tm} \gets [c_p^t, c_p^m, s_p]$
    \State $p_{td} \gets [c_p^t, c_p^d, s_p]$
    \State // four signals: task±method/material
    \State $\{s_j(q,p)\}_{j=1..4} \gets \text{compute\_signals}(q, p_{tm}, p_{td})$
    \State $\widehat{\cos}(q,p) \gets \sum_{j=1..4} \alpha_j \cdot s_j(q,p)$ \quad // $\alpha_j \ge 0, \sum_j \alpha_j = 1$ (learned on dev set)
\EndFor
\State

\State // \textbf{E. Ranking and Return} [\S3.4, Eq.~(10)]
\For{each paper $p \in CP$}
    \State $\text{rank}(q,p) \gets \sum_{i \in \{g,t,m,d\}} w_i \cdot \cos(q, p_i)$ \textbf{OR} $\sum_{j=1..4} \alpha_j \cdot s_j(q,p)$
\EndFor
\State \Return top-$N$ papers by $\text{rank}(q,p)$

\end{algorithmic}
\end{algorithm}

\vspace{1em}
\noindent\textbf{Notes:}
\begin{itemize}
    \item Weight learning: coarse grid coordinate ascent with simplex projection; fixed on test; sensitivity $\pm\{10\%,20\%\}$.
    \item Complexity: cosine over $|CP|=K$ with vector dim $d$ is $O(K \cdot d)$; refinement adds $O(K \cdot d)$; concatenations are $O(d)$.
\end{itemize}

\end{document}